%% file: root.tex
\documentclass[conference]{IEEEtran}
\IEEEoverridecommandlockouts    

\usepackage{multirow}
\usepackage{lipsum}
\usepackage{orcidlink}
\usepackage{amsmath}
\usepackage{amssymb}
\usepackage[ruled,vlined]{algorithm2e}
\usepackage{graphicx}
\graphicspath{{./Figures/}}
\usepackage{amsmath,graphicx,hyperref,amssymb}
\usepackage{tikz}
\usetikzlibrary{shapes.geometric}
\usetikzlibrary{shapes}
\usetikzlibrary{tikzmark, calc,positioning,arrows.meta,quotes}
\usetikzlibrary{calc}
\usetikzlibrary{intersections} 
\usetikzlibrary{backgrounds}
\usetikzlibrary{arrows.meta}
\usepackage{pgfplots}
\usepackage{makecell}
\usepgfplotslibrary{fillbetween}
\usepgfplotslibrary{groupplots}
\pgfplotsset{compat=1.18}
\usepackage{float}
\usepackage{verbatim}
\usepackage{siunitx}
\usepackage{bm}
\usepackage{pgfplots}
\usepackage{xcolor}
\usepackage{dsfont}
\usepackage[normalem]{ulem}
\usepackage{booktabs}
\pgfplotsset{compat=1.18} 
\InputIfFileExists{mathmacros}{}{}
\InputIfFileExists{tu_bs_colors}{}{}

\hypersetup{hidelinks}

\title{\LARGE \bf Doppler RaDAR Single-Frame Anomaly Detection \\and Consistency Assessment \\by Velocity-Conditioned Gaussian Spectral Modeling

\thanks{\textsuperscript{1}Malte Stelzer, Timo Bartels, and Tim Fingscheidt are with the Institute for Communications Technology, Technische Universität Braunschweig, Schleinitzstr. 22, 38106 Braunschweig, Germany \texttt{\{firstname.lastname, t.fingscheidt\}@tu-bs.de}\\
\indent \textsuperscript{2}Dinu Mihailescu-Stoica and Niko Harnge are with Siemens Mobility GmbH, Kiefholzstr. 44, 12435 Berlin, Germany \\
\texttt{\{firstname.lastname\}@siemens.com}\\
\indent \textsuperscript{3} Sören Kohnert, Florian Poprawa, and Marc Zeller are with Siemens Mobility GmbH, Krauss-Maffei-Str. 2, 80997 München\\
\texttt{\{firstname.lastname\}@siemens.com}}
}

 \author{
		Malte Stelzer$^1$\orcidlink{0009-0003-8875-7664}, Timo Bartels$^1$\orcidlink{0009-0002-4917-8849}, Dinu Mihailescu-Stoica$^2$\orcidlink{0009-0009-5698-7303}, Niko Harnge$^2$\orcidlink{0009-0005-6273-0089}, \\ 
Sören Kohnert$^3$\orcidlink{0009-0005-0625-1198}, 
Florian Poprawa$^3$\orcidlink{0009-0002-2920-4640}, Marc Zeller$^3$\orcidlink{0000-0002-6738-7903}, and Tim Fingscheidt$^1$\orcidlink{0000-0002-8895-5041}
}
\begin{document}

\maketitle
\vspace{-0mm}
{\centering\scriptsize
This work has been submitted to the IEEE for possible
publication. Copyright may be transferred without notice,
after which this version may no longer be accessible.\par}
\vspace{+3mm}

\input{sections/0_abstract}

\input{sections/1_Introduction}

\input{sections/3_Method}
\input{sections/4_Experimental_Setup}
\input{sections/5_Experiments}
\input{sections/6_Conclucions}

\bibliographystyle{IEEEtran}
\bibliography{refs}
	
\end{document}

%% file: mathmacros.tex
%
%
%











\newcommand{\putindex}[3]{\vtop{\hbox{\hspace{#3} $#1$}
            \hbox{\raise 6mm \hbox{$\scriptscriptstyle #2$}}}}

\newcommand{\gradx}[0]{\vtop{\hbox{\rm grad}
            \hbox{\raise 2.5mm \hbox{\rm \hspace{2mm} \footnotesize x}}}}

\newcommand{\grady}[0]{\vtop{\hbox{\rm grad}
            \hbox{\raise 2.5mm \hbox{\rm \hspace{2mm} \footnotesize y}}}}

\newcommand{\grad}[1]{\vtop{\hbox{\rm grad}
            \hbox{\raise 2.5mm \hbox{#1}}}}

\newcommand{\btb}{     \begin{tabbing}             }
\newcommand{\bte}{     \end{tabbing}               }

%% file: tu_bs_colors.tex
\definecolor{tu0}{rgb}{0.7451, 0.1176, 0.2353}

\definecolor{tu1}{rgb}{1.0000, 0.8039, 0.0000}
\definecolor{tu11}{rgb}{1.0000, 0.8627, 0.3020}
\definecolor{tu12}{rgb}{1.0000, 0.9020, 0.4980}
\definecolor{tu13}{rgb}{1.0000, 0.9412, 0.6980}
\definecolor{tu14}{rgb}{1.0000, 0.9608, 0.8000}

\definecolor{tu2}{rgb}{0.9804, 0.4314, 0.0000}
\definecolor{tu21}{rgb}{0.9882, 0.6039, 0.3020}
\definecolor{tu22}{rgb}{0.9882, 0.7137, 0.4980}
\definecolor{tu23}{rgb}{0.9922, 0.8275, 0.6980}
\definecolor{tu24}{rgb}{0.9961, 0.8863, 0.8000}

\definecolor{tu3}{rgb}{0.6902, 0.0000, 0.2745}
\definecolor{tu31}{rgb}{0.7529, 0.2000, 0.4196}
\definecolor{tu32}{rgb}{0.8431, 0.4980, 0.6353}
\definecolor{tu33}{rgb}{0.9216, 0.7490, 0.8196}
\definecolor{tu34}{rgb}{0.9529, 0.8510, 0.8902}

\definecolor{tu4}{rgb}{0.4863, 0.8039, 0.9020}
\definecolor{tu41}{rgb}{0.6431, 0.8627, 0.9333}
\definecolor{tu42}{rgb}{0.7412, 0.9020, 0.9490}
\definecolor{tu43}{rgb}{0.8431, 0.9412, 0.9686}
\definecolor{tu44}{rgb}{0.8980, 0.9608, 0.9804}

\definecolor{tu5}{rgb}{0.0000, 0.5020, 0.7059}
\definecolor{tu51}{rgb}{0.3020, 0.6510, 0.7961}
\definecolor{tu52}{rgb}{0.5490, 0.7765, 0.8667}
\definecolor{tu53}{rgb}{0.7490, 0.8745, 0.9255}
\definecolor{tu54}{rgb}{0.8510, 0.9255, 0.9569}

\definecolor{tu6}{rgb}{0.0000, 0.3255, 0.4549}
\definecolor{tu61}{rgb}{0.2510, 0.4941, 0.5922}
\definecolor{tu62}{rgb}{0.5490, 0.6941, 0.7529}
\definecolor{tu63}{rgb}{0.7490, 0.8314, 0.8627}
\definecolor{tu64}{rgb}{0.8510, 0.8980, 0.9176}

\definecolor{tu7}{rgb}{0.7765, 0.9333, 0.0000}
\definecolor{tu71}{rgb}{0.8431, 0.9529, 0.3020}
\definecolor{tu72}{rgb}{0.8863, 0.9647, 0.4980}
\definecolor{tu73}{rgb}{0.9333, 0.9804, 0.6980}
\definecolor{tu74}{rgb}{0.9569, 0.9882, 0.8000}

\definecolor{tu8}{rgb}{0.5373, 0.6431, 0.0000}
\definecolor{tu81}{rgb}{0.6784, 0.7490, 0.3020}
\definecolor{tu82}{rgb}{0.7686, 0.8196, 0.4980}
\definecolor{tu83}{rgb}{0.8588, 0.8941, 0.6980}
\definecolor{tu84}{rgb}{0.9059, 0.9294, 0.8000}

\definecolor{tu9}{rgb}{0.0000, 0.4431, 0.3373}
\definecolor{tu91}{rgb}{0.3020, 0.6118, 0.5373}
\definecolor{tu92}{rgb}{0.5490, 0.7490, 0.7020}
\definecolor{tu93}{rgb}{0.7490, 0.8588, 0.8353}
\definecolor{tu94}{rgb}{0.8549, 0.9176, 0.9059}

\definecolor{tu10}{rgb}{0.8000, 0.0000, 0.6000}
\definecolor{tu101}{rgb}{0.8706, 0.3490, 0.7412}
\definecolor{tu102}{rgb}{0.9216, 0.6000, 0.8392}
\definecolor{tu103}{rgb}{0.9608, 0.8000, 0.9216}
\definecolor{tu104}{rgb}{0.9804, 0.8980, 0.9608}

\definecolor{tu110}{rgb}{0.4627, 0.0000, 0.4627}
\definecolor{tu111}{rgb}{0.5961, 0.2510, 0.5961}
\definecolor{tu112}{rgb}{0.7294, 0.4980, 0.7294}
\definecolor{tu113}{rgb}{0.8392, 0.6980, 0.8392}
\definecolor{tu114}{rgb}{0.9216, 0.8510, 0.9216}

\definecolor{tu120}{rgb}{0.4627, 0.0000, 0.3294}
\definecolor{tu121}{rgb}{0.6118, 0.3020, 0.5333}
\definecolor{tu122}{rgb}{0.7569, 0.5490, 0.6980}
\definecolor{tu123}{rgb}{0.8667, 0.7490, 0.8314}
\definecolor{tu124}{rgb}{0.9216, 0.8510, 0.9020}

\definecolor{tu130}{rgb}{0.0314, 0.0314, 0.0314}
\definecolor{tu131}{rgb}{0.3725, 0.3725, 0.3725}
\definecolor{tu132}{rgb}{0.5882, 0.5882, 0.5882}
\definecolor{tu133}{rgb}{0.7529, 0.7529, 0.7529}
\definecolor{tu134}{rgb}{0.8667, 0.8667, 0.8667}

\definecolor{tu140}{rgb}{0.0000, 0.6875, 0.3125}

%% file: sections/0_abstract.tex
\begin{abstract}
Reliable speed-over-ground estimation is essential for automated railway vehicles. However, the accuracy of a velocity estimate alone does not indicate whether the underlying RaDAR measurement is physically plausible. This paper introduces a velocity-conditioned Gaussian spectral modeling framework for anomaly detection and velocity consistency assessment. A neural encoder estimates the vehicle velocity from complex Doppler spectra, while a Gaussian decoder models the expected spectral distribution conditioned on a velocity estimate. Three complementary inference-time online indicators are derived: the spectral anomaly score (SAS), conditional anomaly score (CAS), and absolute velocity disagreement (AVD). Experiments demonstrate the sensitivity of the scores to velocity deviations. Under controlled spectral anomalies, SAS and CAS outperform conventional spectral indicators in most conditions. Furthermore, AVD exhibits strong correlations with actual velocity estimation errors (Pearson $r>0.99$) for the investigated extra-peak and local anomalies. These results demonstrate the potential of online conditional spectral modeling to assess measurement plausibility and velocity estimation reliability without requiring ground-truth velocities during inference.
\end{abstract}

%% file: sections/1_Introduction.tex
\section{Introduction}
\label{sec:Intro}
Reliable velocity information is essential for railway odometry and the safe operation of increasingly automated rail vehicles: accurate estimates support localization and control, while unreliable measurements can compromise downstream functions \cite{Fingscheidt2022}. Railway vehicles commonly rely on wheel-based odometry, inertial sensors, and satellite navigation, each with limitations under challenging operating conditions \cite{reissland2023highaccuracy, Rahimi2022}. Ground-facing Doppler RaDAR complements these methods with contactless speed-over-ground estimation using Doppler shifts of signals reflected from the track bed \cite{kakuschke2006principles, stelzer2026confidenceawarespeedovergroundestimationdoppler}. \textit{Vehicle velocity} can be inferred from the resulting Doppler spectra using conventional methods such as peak picking, center-of-mass estimation, and spectral cross-correlation \cite{Shariff2017AdaptiveSpectralEstimation}, as well as learning-based approaches \cite{stelzer2026confidenceawarespeedovergroundestimationdoppler}.
However, Doppler RaDAR measurements can be affected by track-bed conditions and multipath propagation \cite{reissland2023highaccuracy}, as well as radar interference \cite{radarinterference2020}, raising the question of how reliably individual measurements can be assessed.

Disturbances in Doppler RaDAR measurements can alter spectral distributions through increased noise, additional peaks, or changes in dominant components. Detecting these deviations helps assess measurement plausibility independently of velocity estimation accuracy. Conventional indicators, including spectral entropy based on Shannon's information measure \cite{shannon1948mathematical} and previously applied to Doppler spectra \cite{Chen2019}, as well as peak-to-noise ratio \cite{meyer2004peaknoise}, characterize selected spectral properties. Signal-based quality assessment has also been investigated for RaDAR-derived velocity measurements \cite{cosoli2012quality,kirincich2012quality}. However, these approaches are tailored to ocean-current measurements and require application-specific processing, limiting their direct applicability to single-frame railway Doppler spectra. Their use of spectral characteristics nevertheless motivates a velocity-conditioned spectral energy (VCSE) baseline considered here, which evaluates spectral energy relative to the predicted velocity. More elaborate Doppler spectral methods include machine-learning-based peak identification \cite{kalesse2019peakfinding} and hierarchical representations of multiple spectral peaks \cite{radenz2019peaktree}.

Beyond RaDAR-specific methods, deep \textit{anomaly detection} includes reconstruction-based, representation-based, and probabilistic approaches that learn characteristics of normal observations \cite{pang2021deepanomaly}. Anomaly detection in time-series data has also been studied \cite{blazquezgarcia2021review}. Recurrent encoder-decoder networks \cite{malhotra2016lstm} and stochastic recurrent models \cite{su2019robust} exploit temporal dependencies to identify anomalies, while related approaches combine recurrent prediction with adaptive thresholds for spacecraft telemetry \cite{hundman2018spacecraft}. Alternatively, deep one-class classification learns representations of normal observations to distinguish anomalous samples \cite{ruff2018deeponeclass}. Unlike these temporal approaches, frame-wise spectral assessment can detect deviations from expected spectral characteristics in individual Doppler measurements without relying on preceding frames.

Beyond detecting anomalous RaDAR measurements, assessing the \textit{reliability} of the resulting \textit{velocity estimates} is equally important. A spectrum may deviate from its expected characteristics without causing a large estimation error, while an erroneous velocity estimate may not be identifiable from simple spectral indicators alone. Research in neural classification considers erroneous predictions and out-of-distribution inputs as distinct but related problems \cite{hendrycks2017baseline}. Learning-based approaches distinguish aleatoric uncertainty arising from observation noise from epistemic uncertainty associated with the learned model \cite{kendall2017uncertainties}, providing a framework for interpreting prediction reliability. However, predictive uncertainty can become less reliable under distribution shifts \cite{ovadia2019uncertainty}. Selective prediction offers another perspective by rejecting predictions when their estimated reliability is insufficient \cite{geifman2019selectivenet}. For railway Doppler RaDAR, a reconstruction-based velocity estimator with an additional confidence head predicting frame-wise error bounds has been introduced  \cite{stelzer2026confidenceawarespeedovergroundestimationdoppler}. However, these estimates do not explicitly assess spectral plausibility or consistency with the predicted velocity, particularly whether a velocity-conditioned spectral model can adequately explain the observation. To address these aspects, we introduce a generative approach based on a velocity-conditioned Gaussian model that learns the expected spectral distribution. This enables assessment of both the best achievable spectral fit and agreement with the encoder-predicted velocity. Comparing the encoder prediction with the velocity favored by the generative model further provides a ground-truth-independent \textit{consistency} indicator that may reveal estimation errors.

\textit{Generative modeling} offers a complementary approach by learning distributions of expected measurements rather than relying solely on handcrafted spectral features. Neural conditional density estimation allows probability distributions, including Gaussian models, to be parameterized by conditioning variables \cite{bishop1994mixturedensity}. Gaussian mixture variational autoencoders have also been investigated for probabilistic anomaly detection in multivariate time series \cite{guo2018gaussianmixture}. Other generative approaches identify anomalies through reconstruction-based models, including adversarial methods \cite{schlegl2019fanogan}. However, high generative likelihood does not necessarily imply that an observation is representative of the training distribution \cite{nalisnick2019deepgenerative}. To address background statistics, likelihood ratios using a separate background model trained on perturbed observations have been proposed \cite{ren2019likelihoodratios}. Direct application to Doppler spectra is less straightforward, since defining perturbations that suppress velocity-related signal structure while preserving background statistics is challenging. This motivates evaluating spectral consistency indicators without a separate background model. Conditioning a probabilistic spectral model on velocity enables evaluation across different candidate velocities. Building on the velocity encoder introduced in \cite{stelzer2026confidenceawarespeedovergroundestimationdoppler}, \textit{this paper proposes a velocity-conditioned Gaussian model of Doppler spectra for anomaly detection and velocity consistency assessment. Three complementary test-time online indicators are derived:} the spectral anomaly score (SAS), measuring the best achievable spectral fit across candidate velocities; the conditional anomaly score (CAS), evaluating spectral plausibility at the encoder-predicted velocity; and the absolute velocity disagreement (AVD), quantifying the difference between the encoder prediction and the velocity preferred by the generative model. We evaluate the sensitivity of learned spectral consistency to velocity deviations, compare the indicators with conventional spectral baselines using controlled synthetic anomalies, and investigate the relationship between AVD and actual velocity estimation errors.

The remainder of this paper is structured as follows. Section \ref{sec:method} presents the method, Section \ref{sec:experimental} the experimental setup, Section \ref{sec:discussion} the results and discussion, and Section \ref{sec:conclusions} the conclusions.

%% file: sections/3_Method.tex
\section{Method}
\label{sec:method}
Fig.~\ref{fig:diagram} summarizes inference: the encoder estimates $\hat v_\ell$ from the Doppler spectrum, while the velocity-conditioned Gaussian decoder predicts the expected spectral mean and variance for plausibility and consistency assessment.

\subsection{Architecture (Inference)}
\textbf{I/Q Data and Preprocessing}: As illustrated by the yellow region in Fig.\,\ref{fig:diagram}, our preprocessing follows \cite{stelzer2026confidenceawarespeedovergroundestimationdoppler}, using the in-phase (I) and quadrature (Q) components of the received Doppler RaDAR measurements, while preserving amplitude and phase information. Let $x(n)\in\mathbb{C}$ denote the resulting complex-valued signal at sample index $n\in\mathbb{Z}$. Using a rectangular window of length $N$, the signal is partitioned into $L$ non-overlapping frames \mbox{$\mathbf{x}_\ell\in\mathbb{C}^{N}$}, indexed by \mbox{$\ell\in\mathcal{L}=\{1,2,\dots,L\}$}. Each frame is transformed into the frequency domain using a $K$-point discrete Fourier transform (DFT), yielding \mbox{$\mathbf{X}_\ell=(X_{\ell,k})\in\mathbb{C}^{K}$}. Here, $K=2048\geq N$, and \mbox{$k\in\mathcal{K}=\{0,1,\dots,K-1\}$} indexes the frequency bins. Frames with $N<K$ are zero-padded before computing the DFT.\newline
\noindent\textbf{Velocity Estimation}: The velocity estimator combines the encoder and velocity \input{figures/architecture_inference}head from \cite{stelzer2026confidenceawarespeedovergroundestimationdoppler}, as shown by the blue region in Fig.~\ref{fig:diagram}. It takes the complex DFT spectrum $\mathbf{X}_\ell$ as input, represented by its real and imaginary components in $\mathbb{R}^{2 \times K}$. Three one-dimensional convolutional blocks progressively reduce the spectral resolution from $K=2048$ to $256$ bins, while increasing the number of feature channels from $2$ to $32$, $64$, and $128$. Each block uses a strided convolution followed by group normalization and ReLU activation. The resulting feature map with dimensions $\mathbb{R}^{128 \times 256}$ is further processed by a convolutional projection and two residual blocks with skip connections, preserving its spectral resolution. Finally, a convolutional prediction head reduces the feature channels from $128$ to $64$ and then to $1$. After dropout and flattening, a fully connected layer maps the remaining $256$ features to the scalar velocity estimate $\hat{v}_\ell \in \mathbb{R}$. For more details, see \cite{stelzer2026confidenceawarespeedovergroundestimationdoppler}. \newline 
\noindent\textbf{Spectral Distribution Estimation}: The green region in Fig.~\ref{fig:diagram} shows the velocity normalization and Gaussian decoder. The estimated velocity $\hat v_\ell$ is normalized to $\hat v_\ell'$ using the training-set mean and standard deviation before entering the decoder. \textit{Inside the decoder,} $\hat v_\ell'$ \textit{is processed in two parallel signal paths.} In the first path, a three-layer perceptron with layer output dimensions $64$, $128$, and $8192$ projects $\hat{v}_\ell'$ into an initial feature map of size $64\times128$. Four interpolation stages progressively increase the spectral resolution to $256$, $512$, $1024$, and finally $2048$ bins. Convolutional layers with SiLU activations \cite{elfwing2018silu} refine the features while reducing the number of channels from $64$ to $32$ and $16$. The resulting feature map is denoted by $\mathbf{f}_\ell\in\mathbb{R}^{16\times2048}$. In the second path, a Gaussian positional profile $\mathbf{g}_\ell\in\mathbb{R}^{1\times2048}$ is constructed at the Doppler position corresponding to the velocity $\hat{v}_\ell$, with a fixed standard deviation of eight bins. This explicitly represents the expected spectral position associated with $\hat{v}_\ell$.

The outputs of both paths are concatenated along the channel dimension, yielding $\mathbf{h}_\ell\in\mathbb{R}^{17\times2048}$. Two final convolutional layers predict the mean $\mu_k(\hat{v}_\ell)$ and log-variance $\log\sigma^2_k(\hat{v}_\ell)$ of the log-magnitude spectrum $\log(|X_{\ell,k}|+\varepsilon)$, with constrained $\log\sigma^2_k(\hat{v}_\ell)\in[-10,4]$.

Thus, for each frequency bin $k\in\mathcal{K}$, the decoder predicts the parameters
\begin{equation}
\left(
\mu_k(\hat{v}_\ell),
\sigma_k^2(\hat{v}_\ell)
\right)\in\mathbb{R}^2
\end{equation}
of a Gaussian distribution conditioned on $\hat{v}_\ell$. Together, these bin-wise distributions form a heteroscedastic \cite{kendall2017uncertainties} diagonal Gaussian model \cite[p.~84]{bishop2006pattern}, assuming conditional independence between frequency bins. The predicted variance accounts for velocity-dependent
variability in the observed log-magnitude spectra.
Consequently, larger predicted variances reduce the
contribution of squared residuals, while the log-variance
term discourages excessively large uncertainty estimates.
\subsection{Anomaly Scores}
The pink region in Fig.~\ref{fig:diagram} illustrates the assessment of spectral plausibility and velocity consistency. A reliable velocity estimate does not necessarily imply an undisturbed RaDAR measurement, while a physically plausible spectrum may still be inconsistent with the estimated velocity. We therefore introduce three \textit{complementary} indicators.

In the following, we derive the conditional anomaly score (CAS) using the mean $\mu_k(\hat v_\ell)$ and variance $\sigma_k^2(\hat v_\ell)$ predicted by the Gaussian decoder in Fig.~\ref{fig:diagram}. Given the log-magnitude observation 
\begin{equation}
    y_{\ell,k}=\log(|X_{\ell,k}|+\varepsilon),
    \label{eq:log_mag}
\end{equation}
where a small $\varepsilon>0$ ensures numerical stability, we define the weighted spectral consistency score (lower is better), inspired by \cite{kendall2017uncertainties}, as
\begin{align}
\label{eq:weighted_nll}
&J_{\ell}(\hat v_\ell)
= \\
&\frac{1}{2\sum_{k\in\mathcal{K}} w_{\ell,k}}
\sum_{k\in\mathcal{K}} w_{\ell,k}
\left(
\frac{(y_{\ell,k}-\mu_k(\hat v_\ell))^2}{\sigma_k^2(\hat v_\ell)}
+\log \sigma_k^2(\hat v_\ell)
\right).\notag
\end{align}
The amplitude-dependent weights are defined as
\begin{align}
\label{eq:amplitude_weights}
&w_{\ell,k}
= \\
&\min\left\{
1+\alpha
\left[
\operatorname{clip}
\left(
\frac{|X_{\ell,k}|}
{\max(q_{0.95}(|\mathbf{X}_{\ell}|),\varepsilon)},
0,1
\right)
\right]^p,
w_{\max}
\right\}, \notag
\end{align}
where $p$ controls the weighting exponent, $w_{\max}$ limits the weights, and $q_{0.95}(\cdot)$ denotes the empirical 95th percentile of the spectral magnitudes:
\begin{equation}
q_{0.95}(|\mathbf{X}_{\ell}|)
= \min \left\{
a \Big|
\frac{1}{|\mathcal{K}|}
\sum_{k\in\mathcal{K}}
\mathds{1} \left(|X_{\ell,k}|\leq a\right)
\geq 0.95
\right\}.
\end{equation}
Here, $\mathds1(\cdot)$ denotes the indicator function, which equals one if the condition is satisfied and zero otherwise. The weighting emphasizes prominent spectral components while retaining contributions from all frequency bins. The Gaussian normalization constant $\frac{1}{2}\log(2\pi)$ is omitted in (\ref{eq:weighted_nll}), as it does not affect score comparisons. Due to the amplitude-dependent weighting, the resulting score is not a normalized negative log-likelihood.

Using (\ref{eq:weighted_nll}), the conditional anomaly score (CAS) directly evaluates spectral consistency at the encoder prediction:
\begin{equation}
\label{eq:conditional_anomaly}
\hat{J}_{\ell}
=
J_{\ell}(\hat{v}_{\ell}).
\end{equation}
A high CAS indicates either a degraded spectrum or inconsistency with the encoder prediction, without distinguishing between these causes.

Next, following the middle path in Fig.~\ref{fig:diagram}, the spectral consistency $J_\ell(v)$ (\ref{eq:weighted_nll}) is evaluated analogously for candidate velocities $v\in\mathcal{V}$, independently of the encoder prediction, which is used for our proposed spectral anomaly score (SAS). The velocity minimizing this score is
\begin{equation}
\label{eq:generative_velocity}
v_{\ell}^{*}
=
\operatorname*{arg\,min}_{v\in\mathcal{V}}
J_{\ell}(v),
\end{equation}
where set $\mathcal{V}$ defines the velocity search space. Note that the $|\mathcal{V}|$ calls of the decoder to deliver parameters $(\mu_k(v),\sigma^2_k(v))$ can be processed offline and the $|\mathcal{V}|\times K\times 2$ parameters be stored in the system. The Gaussian parameters $(\mu_k(v^*_\ell),\sigma^2_k(v^*_\ell))$ corresponding to the consistency-optimal velocity $v_\ell^*$ are then employed in (\ref{eq:weighted_nll}), yielding the spectral anomaly score (SAS):
\begin{equation}
\label{eq:spectral_anomaly}
J^*_{\ell}
=
J_\ell(v_\ell^*)
=
\min_{v\in\mathcal{V}} J_{\ell}(v).
\end{equation}
A high SAS indicates that the measured spectrum is poorly explained even under its most favorable velocity hypothesis. This score is intended to detect spectral \input{figures/architecture_training}anomalies that do not necessarily affect the accuracy of the neural velocity estimate $\hat{v}_\ell$. In particular, a low SAS combined with a high CAS
suggests that the observed spectrum can be explained
by an alternative velocity hypothesis but is inconsistent
with the encoder prediction. A high SAS, in contrast,
indicates poor spectral consistency even after optimizing
over candidate velocities.

Finally, as illustrated by the subtraction and absolute-value operations in Fig.~\ref{fig:diagram}, the absolute velocity disagreement (AVD) compares the encoder prediction with the consistency-optimal velocity:
\begin{equation}
\label{eq:velocity_disagreement}
\Delta v_{\ell}
=
\left|
\hat{v}_{\ell}
-
v_{\ell}^{*}
\right|.
\end{equation}
AVD measures estimator disagreement without requiring ground-truth velocity and must not be interpreted as a direct velocity estimation error.

In summary, SAS assesses spectral plausibility, CAS evaluates consistency at the encoder prediction, and AVD quantifies disagreement between the two velocity hypotheses. Their effectiveness is evaluated \textit{separately} for \textit{spectral degradation} and \textit{velocity estimation} errors. Velocity conditioning is essential: without it, SAS and CAS would coincide, while AVD could not be meaningfully defined.
\subsection{Training}
Our training process is illustrated in Fig.~\ref{fig:training}. \newline
\noindent\textbf{Reference Velocity}: The reference velocity $\overline{v}_\ell$ is obtained following the sensor-fusion procedure described in \cite{stelzer2026confidenceawarespeedovergroundestimationdoppler}, combining measurements from odometrical pulse generators (OPGs) installed on two wheel axles of a rail wagon, a tightly coupled global navigation satellite system (GNSS)/inertial measurement unit (IMU) localization system, and an optical speed correlation sensor. GNSS-derived velocity measurements serve to automatically calibrate the wheel diameters. Limited wheel--rail adhesion during acceleration or braking can cause slip or slide, resulting in discrepancies between wheel rotational speed and actual vehicle speed. Affected frames are identified and excluded by cross-checking the OPG velocity estimates against one another and against the GNSS and optical sensor measurements. Incorporating the optical sensor enables reference-level accuracy of the fused velocity estimate. However, the reference velocity $\overline{v}_\ell$ is obtained through offline post-processing of the recorded data and is therefore not available during online inference. \newline
\noindent\textbf{Losses}: The velocity encoder and Gaussian decoder are trained independently using separate loss functions. For velocity estimation $\hat{v}_\ell$ by the encoder, we minimize the mean absolute error (MAE) across training frames, with the per-frame error defined as
\begin{equation}
J_\ell^\mathrm{MAE}
=
\left|\hat{v}_\ell-\overline{v}_\ell\right|
\label{eq:vel_loss}
\end{equation}
between estimated and reference velocities. For training the Gaussian decoder, we use unperturbed spectra conditioned on reference velocities $\overline{v}_\ell$. The decoder is trained by minimizing the weighted spectral consistency loss 
\begin{equation}
    J_\ell(\overline{v}_\ell),
    \label{eq:loss_gauss}
\end{equation}
defined in (\ref{eq:weighted_nll}) (with $\overline{v}_\ell$ instead of $\hat{v}_\ell$), using the amplitude-dependent weights from (\ref{eq:amplitude_weights}).\newline
\noindent\textbf{Training Noise}: 
Since the velocity estimator is adopted from \cite{stelzer2026confidenceawarespeedovergroundestimationdoppler}, we also employ the frequency-domain noise augmentation strategy used during its training. To increase the robustness of the velocity estimator
against noise-contaminated RaDAR measurements, we
augment the input spectra $\mathbf{X}_{\ell}$ with
artificial additive white complex Gaussian noise during training.
The signal power of each frame $\ell$ is determined
from its spectral coefficients as
\begin{equation}
P^{(X)}_\ell
=
\tfrac{1}{K}
\sum_{k\in\mathcal{K}}
\left|X_{\ell,k}\right|^2
\label{eq:snr_}
\end{equation}
For a specified signal-to-noise ratio ($\mathrm{SNR}$)
in dB, the corresponding noise power is
calculated according to
\begin{equation}
P^{(N)}_\ell
=
\tfrac{P^{(X)}_\ell}
{10^{\mathrm{SNR}/10~\mathrm{dB}}}.
\label{eq:snr}
\end{equation}
Noise samples are independently drawn for each
frequency bin from the complex Gaussian distribution
$N_{\ell,k}\sim
\mathcal{N}_\mathbb{C}(0,P_\ell^{\mathrm{(N)}})$,
where $\mathbf{N}_\ell=(N_{\ell,k}) \in \mathbb{C}^K$.
The augmented spectrum is then obtained as
\begin{equation}
\widetilde{X}_{\ell,k}
=
X_{\ell,k}
+
N_{\ell,k},
\label{eq:noise}
\end{equation}
with $(\tilde{X}_{\ell,k})=\tilde{\mathbf{X}}_\ell$.
As in the synthetic phenomenological anomaly generation described in Section~\ref{sec:experimental}, independent circularly symmetric complex Gaussian noise is added directly to the DFT spectrum across frequency bins rather than to the original I/Q samples. As shown in Fig.~\ref{fig:training}, this augmentation affects only the velocity estimator input, leaving the reference velocity $\overline{v}_\ell$ unchanged. The training SNR values and noise schedule are specified in Section~\ref{sec:experimental}.


%% file: figures/architecture_inference.tex
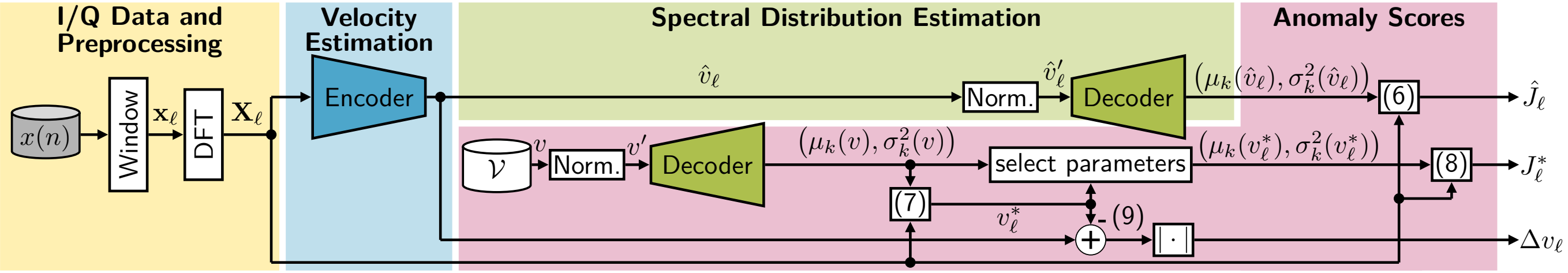
\begin{figure*}[t!]
\centering
\resizebox{\textwidth}{!}{%

        \begin{tikzpicture}[node distance=2cm, auto, font=\sffamily]
            \tikzstyle{arrow} = [-{Triangle[length=4.1pt,width=3.7pt]}, line width=1pt]
            \tikzstyle{line} = [ line width=1pt]
            \tikzstyle{arrow_d} = [-{Triangle[length=4.1pt,width=3.7pt]}, line width=1pt, dashed]
            \tikzstyle{line_d} = [ line width=1pt, dashed]

                \draw [fill=tu13, tu13] (1.7,0.1) rectangle (5.4,-3.45);

                \draw [fill=tu53, tu53] (5.5,0.1) rectangle (7.7,-3.45);

                \draw [fill=tu83, tu83] (7.8,0.1) rectangle (18.1,-1.45);

                \draw [fill=tu33, tu33] (18.2,0.1) rectangle (21.6,-3.45);

                \draw [fill=tu33, tu33] (7.8,-1.55) rectangle (20.6,-3.45);

                \node[cylinder, draw, thick, aspect = 0.8, minimum width = 0.9cm, minimum height = 0.7cm, align=center,  rotate=90, cylinder uses custom fill, cylinder body fill=gray!60!white , cylinder end fill=gray!30!white ] (IQ) at (2.3,-1.7) {};

                \node[anchor=south, align= center] at (2.3,-2) {$x(n)$};
                \node[anchor=south, align= center] at (3.55,-0.40) {\textbf{I/Q Data and}};
                \node[anchor=south, align= left] at (3.55,-0.74) {\textbf{Preprocessing}};

                \node[
                          rectangle,
                          draw,
                          rotate=90,
                          line width=1pt,
                          fill=white,
                          minimum height=0.5cm,
                          minimum width=1.5cm,
                          inner xsep=2pt
                        ] (window) at (3.4,-1.65) {Window};

                \node[
                          rectangle,
                          draw,
                          rotate=90,
                          line width=1pt,
                          fill=white,
                          minimum height=0.5cm,
                          minimum width=1.2cm,
                          inner xsep=2pt
                        ] (dft) at (4.4,-1.65) {DFT};

                \draw[arrow] ([yshift=0.5mm]IQ.south) -- (window.north) ;
                \draw[arrow] (window.south) -- (dft.north) node[midway, above] {$\mathbf{x}_\ell$};
                \draw[line] (dft.south) -- (5.3,-1.65)node[midway, above] {$\mathbf{X}_\ell$};

                \node[anchor=south, align= left] at (6.6,-0.40) {\textbf{Velocity}};
                \node[anchor=south, align= left] at (6.6,-0.66) {\textbf{Estimation}};

\node[
  trapezium,
  trapezium stretches=true,
  trapezium angle=80,
  rotate=270,
  draw,
  line width=1pt,
  fill=tu51,
  minimum height=1.5cm,  
  minimum width=1.1cm,   
  inner sep=2pt
] (enc) at (6.6,-1.15) {\rotatebox{90}{Encoder}};

                \node[anchor=south, align= left] at (12.95,-0.40) {\textbf{Spectral Distribution Estimation}};

\node[
  trapezium,
  trapezium stretches=true,
  trapezium angle=80,
  rotate=90,
  draw,
  line width=1pt,
  fill=tu81,
  minimum height=1.5cm,  
  minimum width=1.1cm,   
  inner sep=2pt
] (dec) at (16.7,-1.15) {\rotatebox{270}{Decoder}};

\node[
          rectangle,
          draw,
          line width=1pt,
          fill=white,
          minimum height=0.3cm,
          minimum width=0.5cm,
          inner xsep=1pt,
          inner ysep=2pt
        ] (norm1) at (15,-1.15) { Norm.};

\node[
          rectangle,
          draw,
          line width=1pt,
          fill=white,
          minimum height=0.3cm,
          minimum width=0.5cm,
          inner xsep=2pt,
          inner ysep=2pt
        ] (norm2) at (16.2,-2.05) {\shortstack{select parameters}};

\draw[arrow] (5.3,-1.15) -- (enc.south);

\node[cylinder, draw, thick, aspect = 0.8, minimum width = 0.9cm, minimum height = 0.7cm, align=center,  rotate=90, cylinder uses custom fill, cylinder body fill=white , cylinder end fill=white ] (velo_V) at (8.3,-2.15) {};

\node[anchor=south, align= center] at (8.3,-2.35) {$\mathcal{V}$};

\node[
          rectangle,
          draw,
          line width=1pt,
          fill=white,
          minimum height=0.3cm,
          minimum width=0.5cm,
          inner xsep=1pt,
          inner ysep=1pt
        ] (inter1) at (13.8,-2.575) { (\ref{eq:generative_velocity}) };

\fill (13.8,-2.05) circle (2pt);
\draw[arrow] (13.8,-2.05) -- (inter1.north);
\fill (13.8,-3.35) circle (2pt);
\draw[arrow] (13.8,-3.35) -- (inter1.south);

\node[
  rectangle,
  draw,
  line width=1pt,
  fill=white,
  minimum height=0.3cm,
  minimum width=0.5cm,
  inner xsep=1pt,
  inner ysep=2pt
] (norm_V) at (9.5,-2.05) { Norm. };

\node[
  trapezium,
  trapezium stretches=true,
  trapezium angle=80,
  rotate=90,
  draw,
  line width=1pt,
  fill=tu81,
  minimum height=1.5cm,
  minimum width=1.1cm,
  inner sep=2pt
] (dec_V) at (11.1,-2.05) {\rotatebox{270}{Decoder}};

\draw[arrow] ([yshift=1mm]velo_V.south) -- (norm_V.west)
  node[midway, above, yshift=0.45mm] {$v$};
\draw[arrow] (norm_V.east) -- (dec_V.north)
  node[midway, above] {$v'$};
\draw[arrow] (dec_V.south) -- (norm2.west) node[midway, above, yshift=-0.5mm, xshift=0mm] {$\left(\mu_{k}(v),\sigma_{k}^{2}(v)\right)$};;

\draw[arrow] (enc.north) -- (norm1.west) node[midway, above, xshift=2mm] {$\hat{v}_\ell$};
\draw[arrow] (norm1.east) -- (dec.north) node[midway, above] {$\hat{v}_\ell'$};

\draw[line] (inter1.east) -- (16.2,-2.575) node[midway, below, yshift=1mm, xshift=-0.2mm] {$v_\ell^*$};
\fill (16.2,-2.575) circle (2pt);
\draw[arrow] (16.2,-2.575) -- (norm2.south);
\draw[arrow] (16.2,-2.575) -- (16.2,-2.85);

\begin{scope}[shift={(16.2,-3.05)}]
    \fill[white] (0,0) circle (2mm);
    \draw[] (0,0) circle (2mm);
    \draw[line width=1pt] (-1.2mm,0) -- (1.2mm,0);
    \draw[line width=1pt] (0,-1.2mm) -- (0,1.2mm);
\end{scope}
\draw[line] (16.3,-2.8) -- (16.4,-2.8);
\node[anchor=south, align= center] at (16.7,-3.05) {(\ref{eq:velocity_disagreement})};
\fill (7.55,-1.15) circle (2pt);
\draw[line] (7.55,-1.15) -- (7.55,-3.05);
\draw[arrow] (7.55,-3.05) -- (16,-3.05);

                \node[
                          rectangle,
                          draw,
                          line width=1pt,
                          fill=white,
                          minimum height=0.3cm,
                          minimum width=0.5cm,
                          inner xsep=1pt,
                          inner ysep=1pt
                        ] (diff1) at (17.3,-3.05) { $|\cdot|$ };



\draw[arrow] (16.4,-3.05) -- (diff1.west);

                \node[
                          rectangle,
                          draw,
                          line width=1pt,
                          fill=white,
                          minimum height=0.3cm,
                          minimum width=0.5cm,
                          inner xsep=1pt,
                          inner ysep=1pt
                        ] (nll1) at (20.3,-1.15) { (\ref{eq:conditional_anomaly}) };

                \node[
                          rectangle,
                          draw,
                          line width=1pt,
                          fill=white,
                          minimum height=0.3cm,
                          minimum width=0.5cm,
                          inner xsep=1pt,
                          inner ysep=1pt
                        ] (nll2) at (21,-2.05) { (\ref{eq:spectral_anomaly}) };

\draw[arrow] (dec.south) -- (nll1.west) node[midway, above, yshift=-1mm] {$\left(\mu_{k}(\hat{v}_\ell),\sigma_{k}^{2}(\hat{v}_\ell)\right)$};

\draw[arrow] (norm2.east) -- (nll2.west) node[midway, above, xshift=-3mm, yshift=-1mm] {$\left(\mu_{k}(v_\ell^*),\sigma_{k}^{2}(v_\ell^*)\right)$};

\draw[arrow] (nll1.east) -- (21.9,-1.15) node[above, xshift=2mm, yshift=-3mm] {$\hat{J}_\ell$};
\draw[arrow] (nll2.east) -- (21.9,-2.05) node[above, xshift=2mm, yshift=-3.6mm] {$J_\ell^*$};

                \node[anchor=south, align= left] at (19.9,-0.40) {\textbf{Anomaly Scores}};

\draw[arrow] (diff1.east) -- (21.9, -3.05)
    node[above, xshift=3mm, yshift=-3mm] {$\Delta v_\ell$};

\fill (5.3,-1.65) circle (2pt);
\draw[line] (5.3,-1.15) -- (5.3, -3.35);
\draw[line] (5.3, -3.35) -- (20.3, -3.35);
\draw[arrow] (20.3, -3.35) -- (nll1.south);
\fill (20.3,-2.5) circle (2pt);
\draw[line] (20.3,-2.5) -- (21,-2.5);
\draw[arrow] (21,-2.5) -- (nll2.south);

        \end{tikzpicture}
}
\vspace*{-2mm}
\caption{\textbf{Inference of our proposed velocity estimation with anomaly detection}. Both decoders are identical.}
\vspace*{-2mm}
\label{fig:diagram}
\end{figure*}

%% file: figures/architecture_training.tex
\begin{figure}[t!]
\centering

        \begin{tikzpicture}[node distance=2cm, auto, font=\sffamily]
            \tikzstyle{arrow} = [-{Triangle[length=4.1pt,width=3.7pt]}, line width=1pt]
            \tikzstyle{line} = [ line width=1pt]
            \tikzstyle{arrow_d} = [-{Triangle[length=4.1pt,width=3.7pt]}, line width=1pt, dashed]
            \tikzstyle{line_d} = [ line width=1pt, dashed]

                \draw [fill=tu13, tu13] (1.7,-0.35) rectangle (5.6,-3.4);

                \draw [fill=tu53, tu53] (5.7,-0.35) rectangle (8.9,-1.8);

                \draw [fill=tu83, tu83] (5.7,-1.9) rectangle (8.9,-3.4);

\node[cylinder, draw, thick, aspect = 0.8, minimum width = 0.9cm, minimum height = 0.7cm, align=center,  rotate=90, cylinder uses custom fill, cylinder body fill=gray!60!white , cylinder end fill=gray!30!white ] (IQ) at (2.3,-2.7) {};

\node[cylinder, draw, thick, aspect = 0.8, minimum width = 0.9cm, minimum height = 0.7cm, align=center,  rotate=90, cylinder uses custom fill, cylinder body fill=gray!60!white , cylinder end fill=gray!30!white ] (IQ) at (2.3,-0.95) {};

                \node[anchor=south, align= center] at (2.3,-3) {$x(n)$};
                \node[anchor=south, align= center] at (2.3,-1.18) {$\overline{v}_\ell$};

                \node[
                          rectangle,
                          rotate=90,
                          draw,
                          line width=1pt,
                          fill=white,
                          minimum height=0.5cm,
                          minimum width=1cm,
                          inner xsep=2pt
                        ] (window) at (3.4,-2.6) {Window};

                \node[
                          rectangle,
                          rotate=90,
                          draw,
                          line width=1pt,
                          fill=white,
                          minimum height=0.5cm,
                          minimum width=1cm,
                          inner xsep=2pt
                        ] (dft) at (4.4,-2.6) {DFT};

\node[
  trapezium,
  trapezium stretches=true,
  trapezium angle=80,
  rotate=270,
  draw,
  line width=1pt,
  fill=tu51,
  minimum height=1.5cm,  
  minimum width=1.1cm,   
  inner sep=2pt
] (enc) at (7.1,-1.15) {\rotatebox{90}{Encoder}};

\node[
  trapezium,
  trapezium stretches=true,
  trapezium angle=80,
  rotate=90,
  draw,
  line width=1pt,
  fill=tu81,
  minimum height=1.5cm,  
  minimum width=1.1cm,   
  inner sep=2pt
] (dec) at (7.1,-2.6) {\rotatebox{270}{Decoder}};

\begin{scope}[shift={(5.2,-1.15)}]
    \fill[white] (0,0) circle (2mm);
    \draw[] (0,0) circle (2mm);
    \draw[line width=1pt] (-1.2mm,0) -- (1.2mm,0);
    \draw[line width=1pt] (0,-1.2mm) -- (0,1.2mm);
\end{scope}

\draw[arrow] (window.south) -- (dft.north) node[midway, below] {$\mathbf{x}_\ell$};
\draw[line] (dft.south) -- (5.2, -2.6) node[midway, below] {$\mathbf{X}_\ell$};
\draw[arrow] (5.2, -2.6) -- (5.2,-1.35);
\fill (5.2, -2.6) circle (2pt);
\draw[line] (5.2, -3.3) -- (5.2, -2.6);
\draw[line] (5.2, -3.3) -- (8.5, -3.3);
\draw[arrow] (5.4, -1.15) -- (enc.south) node[midway, below, xshift=-2mm] {$\tilde{\mathbf{X}}_\ell$};
\draw[line] (6, -0.45) -- (8.5,-0.45) ;
\draw[arrow] (2.75, -2.6) -- (window.north);

\fill (6, -0.85) circle (2pt);
\draw[line] (2.75, -0.85) -- (6, -0.85) node[midway, above, xshift=-3mm, yshift=-0.5mm] {$\overline{v}_\ell$};

\draw[line] (6, -0.45) -- (6, -2.6);
\draw[arrow] (6, -2.6) -- (dec.north);

                \node[
                          rectangle,
                          draw,
                          line width=1pt,
                          fill=white,
                          minimum height=0.3cm,
                          minimum width=0.5cm,
                          inner xsep=1pt,
                          inner ysep=1pt
                        ] (loss1) at (8.5,-1.15) { (\ref{eq:vel_loss}) };

                \node[
                          rectangle,
                          draw,
                          line width=1pt,
                          fill=white,
                          minimum height=0.3cm,
                          minimum width=0.5cm,
                          inner xsep=1pt,
                          inner ysep=1pt
                        ] (loss2) at (8.5,-2.6) { (\ref{eq:loss_gauss}) };

\draw[arrow] (enc.north) -- (loss1.west) node[midway, above] {$\hat{v}_\ell$}; 
\draw[arrow] (dec.south) -- (loss2.west);
\draw[arrow] (8.5,-0.45) -- (loss1.north);
\draw[arrow] (8.5,-3.3) -- (loss2.south);

\draw[arrow] (loss1.east) -- (9.2,-1.15) node[above, xshift=4.7mm, yshift=-3.5mm] {$J_\ell^\mathrm{MAE}$};

\draw[arrow] (loss2.east) -- (9.2,-2.6) node[above, xshift=2mm, yshift=-3.2mm] {$J_\ell$};
\draw[line width=0.5pt] plot[smooth, tension=.7]
coordinates {
    (8.3,-2)
    (8.277,-2.228)
    (8.123,-2.304)
    (8.05,-2.5)
};

\draw[arrow] (4,-1.15) node[above, xshift=-3mm, yshift=-3mm] {$\mathbf{N}_\ell$} -- (5,-1.15) ;

\node[anchor=south, align= center] at (8.3,-2.15) {$\left(\mu_{k}(\overline{v}_\ell),\sigma_{k}^{2}(\overline{v}_\ell)\right)$};

        \end{tikzpicture}
\vspace*{-2mm}
\caption{Independent \textbf{training} of the velocity encoder and Gaussian decoder.}
\vspace*{-2mm}
\label{fig:training}
\end{figure}
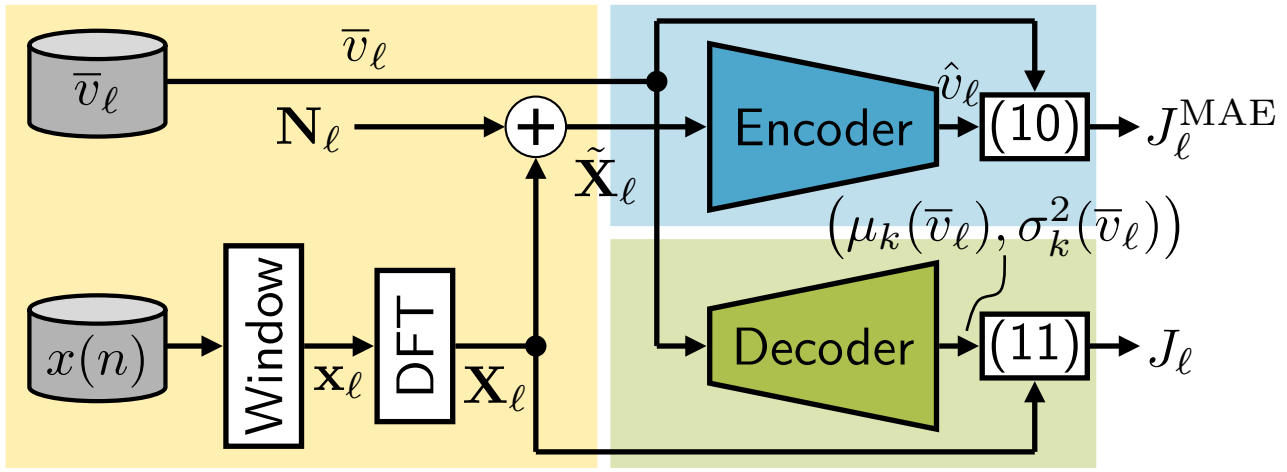

%% file: sections/4_Experimental_Setup.tex
\section{Experimental Setup}
\label{sec:experimental}
\textbf{Dataset}: In railway odometry, Doppler RaDAR complements odometrical pulse generators (OPGs) and inertial sensors by providing contactless ground-relative velocity measurements. The measurement setup follows the configuration described in \cite{stelzer2026confidenceawarespeedovergroundestimationdoppler}. The system employs a continuous-wave (CW) RaDAR with a patch antenna operating at a carrier frequency of $f_c=24$ GHz. To extract the in-phase (I) and quadrature (Q) components, the ground-reflected signal is mixed with two versions of the transmitted signal whose phases differ by 90°. The I/Q imbalance is less than 2\% and is therefore disregarded. Before digitization at $f_s=83.682$ kHz, both channels undergo high-pass filtering to suppress offsets and low-pass filtering to prevent aliasing. The antenna is mounted at an angle of $\vartheta=\ang{37}$ to the horizontal, with a half-power beamwidth of 5°.

The dataset is partitioned into $200\mathrm{k}$ training frames ($\mathcal{D}^\mathrm{train}$), $20\mathrm{k}$ validation frames ($\mathcal{D}^\mathrm{val}$), and $20\mathrm{k}$ test frames ($\mathcal{D}^\mathrm{test}$). Routes are exclusive to individual subsets, which encompass varying weather and ground conditions. The training subset includes 20 routes; the validation and test subsets include 10 routes each. Frame selection aims at matching the velocity distributions across the three subsets as closely as possible. \newline
\noindent\textbf{Signal Augmentation}: Since the dataset contains only measurements under normal operating conditions, we introduce five synthetic anomaly types into real Doppler spectra. These anomalies are motivated by RaDAR signal models and documented effects of multipath propagation, clutter, and interference \cite{richards2022fundamentals,zheng2024ghost,radarinterference2020}, but are intended for controlled evaluation rather than physically accurate simulations of specific RaDAR disturbances. All anomalies are applied independently to the original complex spectra $\mathbf{X}_{\ell}\in\mathbb{C}^{K}$, with $K=2048$, before log-magnitude transformation, leaving reference velocities unchanged. The five anomaly types are introduced exclusively
for evaluation. During training, only the velocity encoder
receives separately sampled Gaussian noise, as described
in Section \ref{sec:method}, whereas the Gaussian decoder is trained
solely on unperturbed spectra.

First, the \textsf{shift} anomaly circularly displaces all complex frequency bins by $\varkappa\in\{4,8,16\}$:
\begin{equation}
 X'_{\ell,k}=X_{\ell,(k-\varkappa)\bmod K}.
\end{equation}
Bins crossing the upper boundary wrap around, preserving spectral energy without renormalization. This follows the DFT frequency-shifting property \cite{oppenheim2010discrete}.

Second, the \textsf{extra peak} anomaly introduces a Gaussian frequency profile with a standard deviation of two bins and a uniformly distributed random common phase. Its center is randomly selected more than 13 bins from the dominant non-DC peak, and its pre-normalization energy is set to a fraction $\eta\in\{0.1,0.3,0.6\}$ of the original total spectral energy. This additional coherent Doppler component is motivated by multipath-induced ghost targets \cite{zheng2024ghost}.

Third, the \textsf{local} anomaly adds independent complex Gaussian samples within a random window of up to nine bins, more than 13 bins from the dominant peak. With pre-normalization energy fractions $\eta\in\{0.1,0.3,0.6\}$, it represents localized noise-like disturbances motivated by clutter and RaDAR interference \cite{richards2022fundamentals,radarinterference2020}.

Fourth, the \textsf{attenuation} anomaly scales the dominant non-DC peak and its four neighboring bins on either side by $\sqrt{1-\eta}$, with $\eta\in\{0.25,0.5,0.9\}$. The removed energy is redistributed as complex Gaussian clutter across valid bins outside this window, reducing peak prominence rather than attenuating the entire spectrum. This is motivated by amplitude variations and multipath effects \cite{richards2022fundamentals,multipath2017analysis}.

By Parseval's theorem, signal energy is proportional to the sum of squared DFT magnitudes \cite{oppenheim2010discrete}. The \textsf{extra peak}, \textsf{local}, and \textsf{attenuation} spectra are normalized to preserve the original energy within valid non-DC bins, isolating spectral changes from total energy variations. The excluded DC region remains unchanged, and the specified energy fractions refer to values before normalization.

Finally, motivated by conventional RaDAR noise models \cite{richards2022fundamentals}, the \textsf{noise} anomaly adds independent circularly symmetric complex Gaussian samples across all bins, including DC. Noise power is derived from the original mean spectral power at SNR levels of $-5$, $0$, $5$, $10$, $15$, $20$, and $30~\mathrm{dB}$. No energy normalization is applied.

All anomalies use reproducible frame- and condition-specific random seeds for controlled evaluation of anomaly detection and encoder–decoder velocity consistency. \newline
\noindent\textbf{Parameterization}: The experiments use frames of length $N=1287$. Noise augmentation follows a curriculum schedule with injection probabilities of 0.3, 0.5, and 0.7 and SNR ranges of $[15,25]$, $[5,20]$, and $[-5,15]$ dB, starting at epochs 11, 26, and 41, respectively. Both networks are trained independently for 80 epochs using AdamW \cite{loshchilov2019decoupled} with a batch size of 128, an initial learning rate of $10^{-3}$, halved every 15 epochs, and weight decay of $10^{-4}$. Encoder dropout is set to 0.1. The log-magnitude transformation (\ref{eq:log_mag}) and amplitude weighting (\ref{eq:amplitude_weights}) use $\varepsilon=10^{-8}$, with $\alpha=5$, $p=1$, and $w_{\max}=10$. SAS and $v_\ell^*$ (\ref{eq:generative_velocity}) are computed by exhaustive grid search over $\mathcal{V}=[-69.4,69.4]$ m/s in 0.1 m/s steps (1,389 candidates). Finer grids yielded negligible improvements.\newline
\noindent\textbf{Reference Methods}:
For comparison, we implement three conventional frequency-domain velocity estimators: peak picking (\texttt{PP}), which selects the dominant spectral bin \cite{richards2010principles}; the center-of-mass approach (\texttt{CMA}), which estimates the Doppler frequency from a weighted spectral distribution \cite{Shariff2017AdaptiveSpectralEstimation}; and cross-correlation (\texttt{XCA}), which matches the measured spectrum against a Gaussian reference profile \cite{Shariff2017AdaptiveSpectralEstimation}. For \texttt{PP}, seven bins around DC are excluded. A detailed description and comparative evaluation of these methods are provided in \cite{stelzer2026confidenceawarespeedovergroundestimationdoppler}, which also introduces the velocity estimation approach (encoder) employed in this work.

For spectral anomaly detection, we consider three
conventional signal characteristics that do not
require a learned spectral model.
First, spectral entropy (\texttt{SE}), derived from Shannon's information measure \cite{shannon1948mathematical}, quantifies the dispersion of normalized spectral energy across frequency bins. Higher values indicate a more uniform energy distribution, which can arise from additional spectral components or an elevated noise floor.
Second, the peak-to-noise ratio (\texttt{PNR}) \cite{meyer2004peaknoise} measures the
prominence of the dominant spectral component
relative to the estimated spectral noise floor.
A reduction in \texttt{PNR} can indicate attenuation
or increased background noise.
Signal-based quality indicators have previously been investigated for assessing the reliability of RaDAR-derived velocity measurements \cite{cosoli2012quality,kirincich2012quality}. Motivated by this principle, we define a simple velocity-conditioned spectral energy (VCSE) baseline that measures the fraction of non-DC spectral power outside a window of $\pm 4$ DFT bins around the Doppler position predicted by the velocity encoder. Higher scores indicate less spectral energy near the expected Doppler position and thus greater disagreement between the measured spectrum and the estimated velocity.

%% file: sections/5_Experiments.tex
\section{Experimental Validation and Discussion}
\label{sec:discussion}
\input{figures/results}
To verify the velocity encoder reimplementation from \cite{stelzer2026confidenceawarespeedovergroundestimationdoppler}, Fig.~\ref{fig:velocity} compares its MAE (after (\ref{eq:vel_loss})) with conventional estimators under varying noise conditions. For each method and SNR, the MAE is computed by averaging the absolute differences between the estimated and reference velocities $\overline{v}_\ell$ over all samples in $\mathcal{D}^{\mathrm{test}}$, i.e., $|\hat{v}_\ell-\overline{v}_\ell|$ for the reimplemented encoder. The reimplemented encoder \cite{stelzer2026confidenceawarespeedovergroundestimationdoppler} achieves an MAE of approximately 0.15--0.35 m/s across the evaluated SNRs, outperforming all conventional methods. \textit{These results are consistent with those in} \cite{stelzer2026confidenceawarespeedovergroundestimationdoppler} \textit{and validate our encoder reimplementation.}

Having verified the velocity encoder, we next conduct a separate sensitivity test to examine how the learned spectral model responds to deviations from the reference velocity $\overline{v}_\ell$. Fig.~\ref{fig:nll_velocity_deviation} shows the mean of the proposed spectral consistency score employing (\ref{eq:weighted_nll}) as a function of $\Delta\tilde{v}_\ell=\overline{v}_\ell-\tilde{v}_\ell$, where $\tilde{v}_\ell$ is a test velocity supplied to normalization and decoding. Results are averaged over frames grouped by ground-truth velocity interval. The dashed line indicates the mean of the proposed spectral anomaly score $J_\ell^*$ (\ref{eq:spectral_anomaly}), which is independent of the supplied test velocity. Across all intervals, the consistency exhibits a minimum near zero deviation and increases as the supplied velocity moves away from the ground truth. The flatter response for velocities in $[-10,10)$ m/s may result from Doppler components being difficult to distinguish from the DC peak when their frequencies are close to or overlap with it, reducing the score's sensitivity to velocity deviations. This may be particularly relevant on the positive-deviation side, where the evaluated Doppler position can approach the DC region. \textit{Overall, the results indicate that the Gaussian decoder captures the relationship between Doppler spectra and velocity, providing a basis for anomaly detection and velocity consistency assessment.}

\input{figures/results_nll_deltav}To assess whether the proposed consistency scores provide an advantage over conventional spectral indicators, Tab.~\ref{tab:perturbation_auroc} compares their anomaly detection performance with spectral entropy (\texttt{SE}) \cite{shannon1948mathematical}, (negative) peak-to-noise ratio (\texttt{PNR}) \cite{meyer2004peaknoise}, and velocity-conditioned spectral energy (\texttt{VCSE}). Controlled synthetic anomalies of varying strengths are applied to the measured Doppler spectra to evaluate each method's ability to distinguish anomalous from unmodified measurements. Detection performance is quantified using the area under the receiver operating characteristic curve (AUROC), which assesses discrimination rather than measurement rejection; the latter would require additional error quantification and calibrated decision criteria.\newline
The SAS $J_\ell^*$ (\ref{eq:spectral_anomaly}) and CAS $\hat J_\ell$ (\ref{eq:conditional_anomaly}) achieve the highest AUROC values for most anomalies. For spectral \textsf{shifts}, they reach 0.752...0.930 and 0.759...0.934, respectively, whereas the conventional indicators remain at or below 0.692. For \textsf{extra peak} and \textsf{local} anomalies, both scores achieve AUROC values between 0.965 and 0.994, outperforming spectral entropy (\texttt{SE}) \cite{shannon1948mathematical} and \texttt{negative PNR} \cite{meyer2004peaknoise}. The slightly higher AUROC of CAS in several conditions may result from its sensitivity to inconsistencies at the encoder-predicted velocity, whereas SAS minimizes the spectral mismatch over candidate velocities. \texttt{VCSE} remains competitive for stronger anomalies, reaching up to 0.992. \textsf{Attenuation} and \textsf{noise} are detected reliably by several methods, although the consistency scores retain a clear advantage at intermediate SNRs, e.g., achieving 0.984...0.985 at 20 dB compared with 0.947 for negative PNR \cite{meyer2004peaknoise}. At 30 dB, however, detection performance approaches chance level for several methods. The AVD $\Delta v_\ell$ (\ref{eq:velocity_disagreement}) also detects \textsf{extra peak} and \textsf{local} anomalies effectively, but is generally less sensitive than the consistency scores. \textit{Overall, the results demonstrate that the learned spectral consistency (CAS) provides an effective anomaly indicator across the evaluated synthetic conditions. The complementary velocity disagreement (AVD) additionally offers a measure of consistency between the encoder prediction and the velocity favored by the generative model.}

\input{figures/results_auroc}\input{figures/results_scatter} While the preceding results demonstrate the anomaly detection capability of the proposed scores, the AVD $\Delta v_\ell$ (\ref{eq:velocity_disagreement}) may also provide information about individual velocity estimation errors. Since ground-truth velocities are generally unavailable during inference, Fig.~\ref{fig:velocity_scatter} examines the relationship between our AVD $\Delta v_\ell$ (\ref{eq:velocity_disagreement}) and the actual estimation error $|\hat{v}_\ell-\overline{v}_\ell|$ (\ref{eq:vel_loss}) for \textsf{extra peak} and \textsf{local} anomalies at strength $\eta=0.6$. The scatter plot shows 1,000 randomly selected samples per anomaly from $\mathcal{D}^{\mathrm{test}}$, while the statistics are computed using all 20k samples per anomaly. Both anomalies exhibit strong positive correlations, with Pearson $r=0.993$ and $r=0.995$ and Spearman $\rho=0.987$ and $\rho=0.995$, respectively. Most samples lie close to the identity line, including less frequent samples with substantially larger estimation errors. Moreover, AVD deviates from the actual encoder error by at most 1 m/s for over 95\% of samples in both conditions. This agreement is explained by the consistency-optimal velocity $v_\ell^*$ remaining substantially more accurate than the encoder estimate $\hat v_\ell$ under these anomalies: its MAE is 0.459 and 0.455 m/s, compared with 12.547 and 19.413 m/s for $\hat v_\ell$, respectively. Conversely, for undisturbed measurements, the encoder achieves a lower MAE (0.152 versus 0.339 m/s). \textit{Thus, the AVD $\Delta v_\ell$} (\ref{eq:velocity_disagreement}) \textit{can closely approximate the actual encoder error without requiring ground-truth information during inference under the investigated anomalies.} These results establish sensitivity to controlled synthetic anomalies in real RaDAR spectra, while performance under naturally occurring disturbances remains to be investigated. Since the Gaussian decoder models spectral log-magnitudes,
phase-only disturbances cannot be detected directly by
SAS. Such disturbances may nevertheless affect CAS or
AVD if they alter the encoder's velocity prediction. Future work will address more physically realistic CW-RaDAR anomaly modeling and validation under real disturbances.

%% file: figures/results.tex
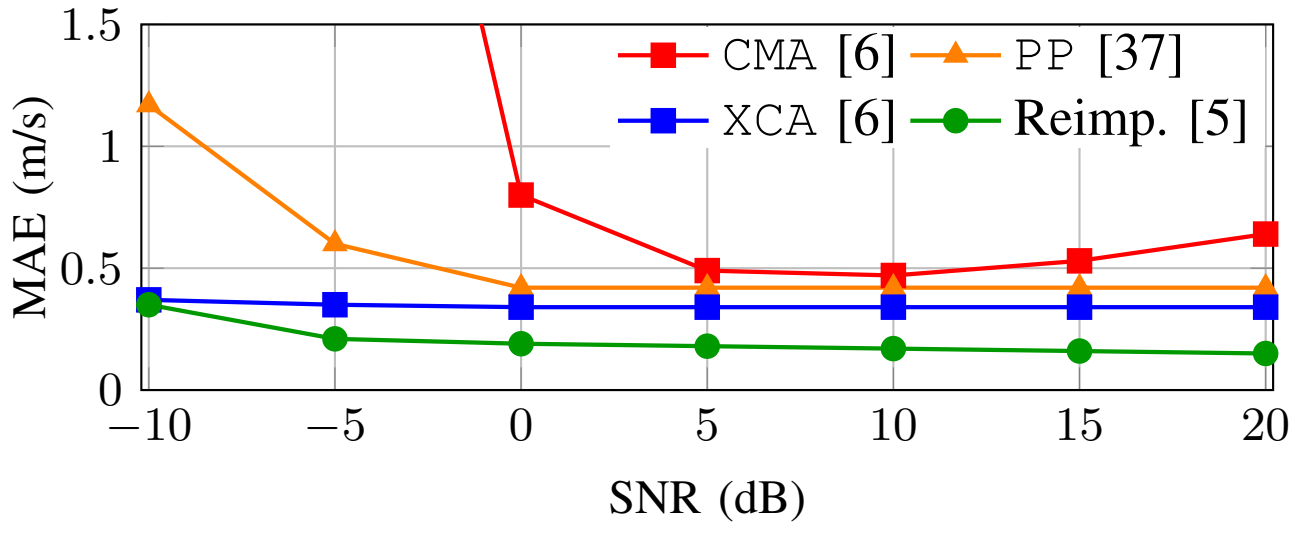
\begin{figure}[h]
    \centering
\begin{tikzpicture}

\begin{axis}[
    name=mainaxis,
    width=0.5\textwidth,
    height=4cm,
    xlabel={SNR (dB)},
    ylabel={MAE (m/s)},
    ylabel style={yshift=-2mm},
    xmin=-10.2, xmax=20.2,
    ymin=0, ymax=1.5,
    grid=both,
    legend pos=north east,
    legend cell align=left,
    legend style={draw=none, fill=white,xshift=1.8mm, legend columns=2,
    yshift=0.57mm,inner sep=1pt, cells={align=left},  inner xsep=1pt,
    inner ysep=-1pt},
    tick label style={font=\small},
    label style={font=\small},
]

\addplot[
    thick,
    draw=red,
    mark=square*,
    mark options={
        draw=red,
        fill=red
    },
    error bars/.cd,
        y dir=both,
        y explicit,
        error bar style={draw=red}
] coordinates {
    (-10,25.07)
    (-5,4.1)
    (0,0.8)  
    (5,0.49)  
    (10,0.47)  
    (15,0.53) 
    (20,0.64)
};
\addlegendentry{\texttt{CMA} \cite{Shariff2017AdaptiveSpectralEstimation}}

\addplot[
    thick,
    draw=orange,
    mark=triangle*,
    mark options={
        draw=orange,
        fill=orange
    },
    error bars/.cd,
        y dir=both,
        y explicit,
        error bar style={draw=orange}
] coordinates {
    (-10,1.17)
    (-5,0.6) 
    (0,0.42)   
    (5,0.42)  
    (10,0.42) 
    (15,0.42)  
    (20,0.42) 
};
\addlegendentry{\texttt{PP} \cite{richards2010principles}}

\addplot[
    thick,
    draw=blue,
    mark=square*,
    mark options={
        draw=blue,
        fill=blue
    },
    error bars/.cd,
        y dir=both,
        y explicit,
        error bar style={draw=blue}
] coordinates {
    (-10,0.37)
    (-5,0.35)
    (0,0.34)  
    (5,0.34)  
    (10,0.34)  
    (15,0.34) 
    (20,0.34)
};
\addlegendentry{\texttt{XCA} \cite{Shariff2017AdaptiveSpectralEstimation}}

\addplot[
    thick,
    draw=green!60!black,
    mark=*,
    mark options={
        draw=green!60!black,
        fill=green!60!black
    },
    error bars/.cd,
        y dir=both,
        y explicit,
        error bar style={draw=green!60!black}
] coordinates {
    (-10,0.35)
    (-5,0.21) 
    (0,0.19)   
    (5,0.18)   
    (10,0.17)  
    (15,0.16)  
    (20,0.15)
};
\addlegendentry{Reimp. \cite{stelzer2026confidenceawarespeedovergroundestimationdoppler}}

\end{axis}
\end{tikzpicture}\vspace*{-2mm}
    \caption{\textbf{Velocity MAE} (computed using (\ref{eq:vel_loss})) on noisy $\mathcal{D}^\mathrm{test}$.}
    \label{fig:velocity}
    \vspace*{-2mm}
\end{figure}

%% file: figures/results_nll_deltav.tex
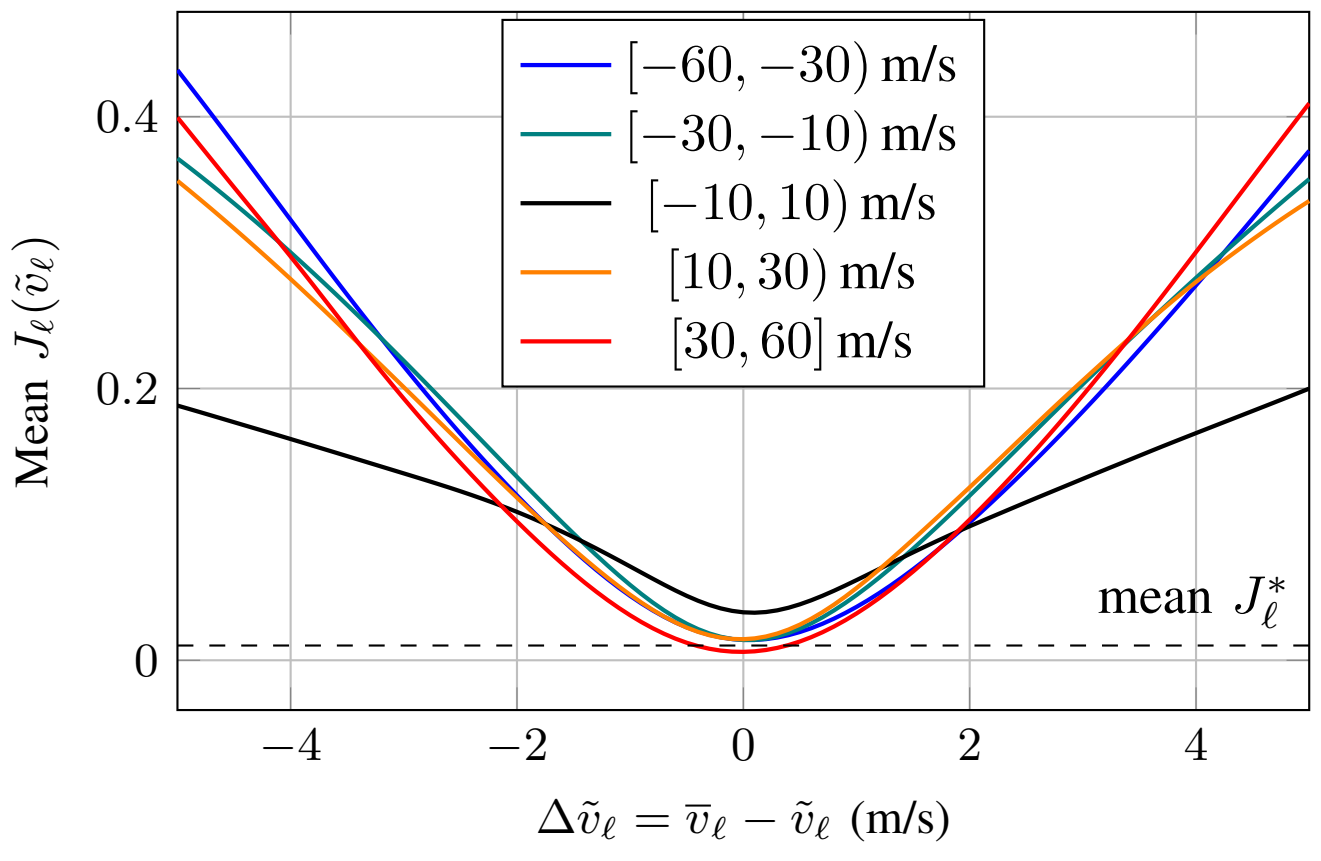
\begin{figure}[t]
\centering
\begin{tikzpicture}
\begin{axis}[
width=0.5\textwidth,
height=0.70\columnwidth,
xlabel={$\Delta \tilde{v}_\ell=\overline{v}_\ell-\tilde{v}_\ell$ (m/s)},
ylabel={Mean $J_\ell(\tilde{v}_\ell)$},
xmin=-5,xmax=5,
grid=major,
tick label style={font=\small},
label style={font=\small},
legend style={
    at={(0.5,0.99)},
    anchor=north,
    fill=white,
    fill opacity=1,
    draw=black
}
]
\addplot[blue,thick,no marks] coordinates {
(-5.0,0.434355987)
(-4.9,0.423609179)
(-4.8,0.412780628)
(-4.7,0.401878318)
(-4.6,0.390907388)
(-4.5,0.379873635)
(-4.4,0.368782600)
(-4.3,0.357642381)
(-4.2,0.346466598)
(-4.1,0.335273171)
(-4.0,0.324086104)
(-3.9,0.312929418)
(-3.8,0.301824429)
(-3.7,0.290789342)
(-3.6,0.279836695)
(-3.5,0.268976697)
(-3.4,0.258215581)
(-3.3,0.247557552)
(-3.2,0.237006706)
(-3.1,0.226565509)
(-3.0,0.216239549)
(-2.9,0.206036916)
(-2.8,0.195968843)
(-2.7,0.186048119)
(-2.6,0.176284221)
(-2.5,0.166685548)
(-2.4,0.157257499)
(-2.3,0.148004144)
(-2.2,0.138929707)
(-2.1,0.130037148)
(-2.0,0.121331287)
(-1.9,0.112817410)
(-1.8,0.104502881)
(-1.7,0.096399199)
(-1.6,0.088521294)
(-1.5,0.080889998)
(-1.4,0.073529306)
(-1.3,0.066465910)
(-1.2,0.059728874)
(-1.1,0.053347290)
(-1.0,0.047351717)
(-0.9,0.041772409)
(-0.8,0.036639937)
(-0.7,0.031985390)
(-0.6,0.027838563)
(-0.5,0.024228328)
(-0.4,0.021180469)
(-0.3,0.018717124)
(-0.2,0.016855866)
(-0.1,0.015608130)
(0.0,0.014978905)
(0.1,0.014965860)
(0.2,0.015559922)
(0.3,0.016745684)
(0.4,0.018501649)
(0.5,0.020799914)
(0.6,0.023607939)
(0.7,0.026891383)
(0.8,0.030616773)
(0.9,0.034754301)
(1.0,0.039277443)
(1.1,0.044163900)
(1.2,0.049394450)
(1.3,0.054951055)
(1.4,0.060816786)
(1.5,0.066973374)
(1.6,0.073402301)
(1.7,0.080084394)
(1.8,0.086999948)
(1.9,0.094131702)
(2.0,0.101463322)
(2.1,0.108980660)
(2.2,0.116669925)
(2.3,0.124516575)
(2.4,0.132508181)
(2.5,0.140633711)
(2.6,0.148886680)
(2.7,0.157264114)
(2.8,0.165764238)
(2.9,0.174386760)
(3.0,0.183128940)
(3.1,0.191987280)
(3.2,0.200957115)
(3.3,0.210032063)
(3.4,0.219206368)
(3.5,0.228472287)
(3.6,0.237822780)
(3.7,0.247251395)
(3.8,0.256752346)
(3.9,0.266323736)
(4.0,0.275964533)
(4.1,0.285674931)
(4.2,0.295452384)
(4.3,0.305288884)
(4.4,0.315174434)
(4.5,0.325097309)
(4.6,0.335048628)
(4.7,0.345020241)
(4.8,0.355000697)
(4.9,0.364976313)
(5.0,0.374929496)
};
\addlegendentry{$[-60,-30)$\,m/s}
\addplot[teal,thick,no marks] coordinates {
(-5.0,0.369321547)
(-4.9,0.362980432)
(-4.8,0.356479819)
(-4.7,0.349840164)
(-4.6,0.343082783)
(-4.5,0.336219304)
(-4.4,0.329249555)
(-4.3,0.322159776)
(-4.2,0.314941236)
(-4.1,0.307599842)
(-4.0,0.300148156)
(-3.9,0.292602696)
(-3.8,0.284971148)
(-3.7,0.277239939)
(-3.6,0.269388365)
(-3.5,0.261418575)
(-3.4,0.253350739)
(-3.3,0.245205482)
(-3.2,0.237000178)
(-3.1,0.228737918)
(-3.0,0.220406406)
(-2.9,0.211998302)
(-2.8,0.203522145)
(-2.7,0.194997067)
(-2.6,0.186447101)
(-2.5,0.177886727)
(-2.4,0.169312077)
(-2.3,0.160717270)
(-2.2,0.152112937)
(-2.1,0.143521465)
(-2.0,0.134966463)
(-1.9,0.126466129)
(-1.8,0.118030669)
(-1.7,0.109668720)
(-1.6,0.101396893)
(-1.5,0.093241205)
(-1.4,0.085233170)
(-1.3,0.077408137)
(-1.2,0.069803336)
(-1.1,0.062457411)
(-1.0,0.055415960)
(-0.9,0.048733989)
(-0.8,0.042472249)
(-0.7,0.036694511)
(-0.6,0.031466321)
(-0.5,0.026852694)
(-0.4,0.022915644)
(-0.3,0.019711304)
(-0.2,0.017285847)
(-0.1,0.015673227)
(0.0,0.014894510)
(0.1,0.014957271)
(0.2,0.015855001)
(0.3,0.017566329)
(0.4,0.020054659)
(0.5,0.023269528)
(0.6,0.027149927)
(0.7,0.031628488)
(0.8,0.036636559)
(0.9,0.042110109)
(1.0,0.047993475)
(1.1,0.054240564)
(1.2,0.060812838)
(1.3,0.067674996)
(1.4,0.074792151)
(1.5,0.082129394)
(1.6,0.089654713)
(1.7,0.097341343)
(1.8,0.105167684)
(1.9,0.113114741)
(2.0,0.121162012)
(2.1,0.129287028)
(2.2,0.137468309)
(2.3,0.145688809)
(2.4,0.153936187)
(2.5,0.162198236)
(2.6,0.170459683)
(2.7,0.178701726)
(2.8,0.186905970)
(2.9,0.195060235)
(3.0,0.203160097)
(3.1,0.211207384)
(3.2,0.219205694)
(3.3,0.227155907)
(3.4,0.235056273)
(3.5,0.242905306)
(3.6,0.250705225)
(3.7,0.258460843)
(3.8,0.266174846)
(3.9,0.273843985)
(4.0,0.281458184)
(4.1,0.289007163)
(4.2,0.296487266)
(4.3,0.303902786)
(4.4,0.311260508)
(4.5,0.318563184)
(4.6,0.325806632)
(4.7,0.332983812)
(4.8,0.340091106)
(4.9,0.347129729)
(5.0,0.354102094)
};
\addlegendentry{$[-30,-10)$\,m/s}
\addplot[black,thick,no marks] coordinates {
(-5.0,0.187418130)
(-4.9,0.185028328)
(-4.8,0.182598681)
(-4.7,0.180138106)
(-4.6,0.177664838)
(-4.5,0.175192945)
(-4.4,0.172727377)
(-4.3,0.170267901)
(-4.2,0.167803610)
(-4.1,0.165318931)
(-4.0,0.162815269)
(-3.9,0.160306106)
(-3.8,0.157796020)
(-3.7,0.155281736)
(-3.6,0.152758049)
(-3.5,0.150220152)
(-3.4,0.147674893)
(-3.3,0.145131791)
(-3.2,0.142591083)
(-3.1,0.140052263)
(-3.0,0.137512930)
(-2.9,0.134960298)
(-2.8,0.132384523)
(-2.7,0.129780010)
(-2.6,0.127124453)
(-2.5,0.124389686)
(-2.4,0.121563007)
(-2.3,0.118638998)
(-2.2,0.115614648)
(-2.1,0.112477648)
(-2.0,0.109197917)
(-1.9,0.105753056)
(-1.8,0.102153688)
(-1.7,0.098427042)
(-1.6,0.094585278)
(-1.5,0.090622172)
(-1.4,0.086527818)
(-1.3,0.082300658)
(-1.2,0.077955484)
(-1.1,0.073518619)
(-1.0,0.069019002)
(-0.9,0.064494405)
(-0.8,0.059996489)
(-0.7,0.055589634)
(-0.6,0.051354195)
(-0.5,0.047387979)
(-0.4,0.043799617)
(-0.3,0.040699186)
(-0.2,0.038190871)
(-0.1,0.036364879)
(0.0,0.035289363)
(0.1,0.035002372)
(0.2,0.035503101)
(0.3,0.036751559)
(0.4,0.038675522)
(0.5,0.041177823)
(0.6,0.044149446)
(0.7,0.047491425)
(0.8,0.051126593)
(0.9,0.054985024)
(1.0,0.058988842)
(1.1,0.063064814)
(1.2,0.067163564)
(1.3,0.071264514)
(1.4,0.075357322)
(1.5,0.079418943)
(1.6,0.083419394)
(1.7,0.087342293)
(1.8,0.091187618)
(1.9,0.094952998)
(2.0,0.098630084)
(2.1,0.102223249)
(2.2,0.105760205)
(2.3,0.109279021)
(2.4,0.112799833)
(2.5,0.116310622)
(2.6,0.119787879)
(2.7,0.123230584)
(2.8,0.126661663)
(2.9,0.130102007)
(3.0,0.133550299)
(3.1,0.136983105)
(3.2,0.140381557)
(3.3,0.143758769)
(3.4,0.147144807)
(3.5,0.150548927)
(3.6,0.153946159)
(3.7,0.157305202)
(3.8,0.160619824)
(3.9,0.163905966)
(4.0,0.167178847)
(4.1,0.170441395)
(4.2,0.173694568)
(4.3,0.176946467)
(4.4,0.180202786)
(4.5,0.183462699)
(4.6,0.186724154)
(4.7,0.189992103)
(4.8,0.193280899)
(4.9,0.196598508)
(5.0,0.199930978)
};
\addlegendentry{$[-10,10)$\,m/s}
\addplot[orange,thick,no marks] coordinates {
(-5.0,0.352855634)
(-4.9,0.346041313)
(-4.8,0.339129210)
(-4.7,0.332119860)
(-4.6,0.325014743)
(-4.5,0.317817359)
(-4.4,0.310532181)
(-4.3,0.303161639)
(-4.2,0.295705786)
(-4.1,0.288164053)
(-4.0,0.280537174)
(-3.9,0.272828470)
(-3.8,0.265042482)
(-3.7,0.257186087)
(-3.6,0.249268435)
(-3.5,0.241298189)
(-3.4,0.233282762)
(-3.3,0.225230267)
(-3.2,0.217149062)
(-3.1,0.209045659)
(-3.0,0.200923302)
(-2.9,0.192781223)
(-2.8,0.184618770)
(-2.7,0.176439267)
(-2.6,0.168250205)
(-2.5,0.160063784)
(-2.4,0.151894669)
(-2.3,0.143757734)
(-2.2,0.135666987)
(-2.1,0.127635902)
(-2.0,0.119680627)
(-1.9,0.111818988)
(-1.8,0.104068209)
(-1.7,0.096444285)
(-1.6,0.088961526)
(-1.5,0.081634079)
(-1.4,0.074479030)
(-1.3,0.067520508)
(-1.2,0.060791418)
(-1.1,0.054332432)
(-1.0,0.048190859)
(-0.9,0.042418105)
(-0.8,0.037066933)
(-0.7,0.032188957)
(-0.6,0.027834935)
(-0.5,0.024056658)
(-0.4,0.020906352)
(-0.3,0.018434274)
(-0.2,0.016685665)
(-0.1,0.015697893)
(0.0,0.015497743)
(0.1,0.016098579)
(0.2,0.017498835)
(0.3,0.019681211)
(0.4,0.022613426)
(0.5,0.026252046)
(0.6,0.030546767)
(0.7,0.035442476)
(0.8,0.040880992)
(0.9,0.046801508)
(1.0,0.053139383)
(1.1,0.059829276)
(1.2,0.066809263)
(1.3,0.074020351)
(1.4,0.081405149)
(1.5,0.088910848)
(1.6,0.096494651)
(1.7,0.104129250)
(1.8,0.111805320)
(1.9,0.119526690)
(2.0,0.127305083)
(2.1,0.135152811)
(2.2,0.143068370)
(2.3,0.151032415)
(2.4,0.159017077)
(2.5,0.166992794)
(2.6,0.174931740)
(2.7,0.182809405)
(2.8,0.190608943)
(2.9,0.198330446)
(3.0,0.205986133)
(3.1,0.213588406)
(3.2,0.221145799)
(3.3,0.228658717)
(3.4,0.236115701)
(3.5,0.243493472)
(3.6,0.250764160)
(3.7,0.257904607)
(3.8,0.264901554)
(3.9,0.271748149)
(4.0,0.278436734)
(4.1,0.284962582)
(4.2,0.291335089)
(4.3,0.297574389)
(4.4,0.303698455)
(4.5,0.309716933)
(4.6,0.315632126)
(4.7,0.321437847)
(4.8,0.327117660)
(4.9,0.332651293)
(5.0,0.338022739)
};
\addlegendentry{$[10,30)$\,m/s}
\addplot[red,thick,no marks] coordinates {
(-5.0,0.399470046)
(-4.9,0.389209660)
(-4.8,0.378906532)
(-4.7,0.368589807)
(-4.6,0.358298311)
(-4.5,0.348057973)
(-4.4,0.337871301)
(-4.3,0.327724423)
(-4.2,0.317583618)
(-4.1,0.307394722)
(-4.0,0.297102472)
(-3.9,0.286670782)
(-3.8,0.276100199)
(-3.7,0.265429721)
(-3.6,0.254719835)
(-3.5,0.244037607)
(-3.4,0.233440251)
(-3.3,0.222970273)
(-3.2,0.212656329)
(-3.1,0.202513159)
(-3.0,0.192547698)
(-2.9,0.182759429)
(-2.8,0.173144334)
(-2.7,0.163698172)
(-2.6,0.154415992)
(-2.5,0.145295615)
(-2.4,0.136335520)
(-2.3,0.127536003)
(-2.2,0.118900497)
(-2.1,0.110433962)
(-2.0,0.102145911)
(-1.9,0.094048413)
(-1.8,0.086156763)
(-1.7,0.078490136)
(-1.6,0.071068958)
(-1.5,0.063917041)
(-1.4,0.057059004)
(-1.3,0.050520648)
(-1.2,0.044328855)
(-1.1,0.038509662)
(-1.0,0.033089600)
(-0.9,0.028094347)
(-0.8,0.023549673)
(-0.7,0.019482151)
(-0.6,0.015917689)
(-0.5,0.012881275)
(-0.4,0.010394187)
(-0.3,0.008472922)
(-0.2,0.007128671)
(-0.1,0.006367047)
(0.0,0.006189088)
(0.1,0.006591709)
(0.2,0.007568544)
(0.3,0.009110248)
(0.4,0.011205247)
(0.5,0.013839466)
(0.6,0.016996679)
(0.7,0.020658064)
(0.8,0.024801913)
(0.9,0.029405349)
(1.0,0.034444235)
(1.1,0.039894740)
(1.2,0.045732731)
(1.3,0.051932795)
(1.4,0.058470103)
(1.5,0.065319643)
(1.6,0.072458937)
(1.7,0.079868141)
(1.8,0.087529136)
(1.9,0.095426515)
(2.0,0.103544922)
(2.1,0.111871493)
(2.2,0.120395478)
(2.3,0.129108197)
(2.4,0.138004827)
(2.5,0.147081347)
(2.6,0.156335246)
(2.7,0.165763061)
(2.8,0.175358729)
(2.9,0.185115382)
(3.0,0.195023064)
(3.1,0.205072131)
(3.2,0.215253436)
(3.3,0.225558498)
(3.4,0.235982137)
(3.5,0.246517427)
(3.6,0.257155330)
(3.7,0.267882261)
(3.8,0.278681774)
(3.9,0.289540706)
(4.0,0.300447544)
(4.1,0.311393417)
(4.2,0.322368801)
(4.3,0.333362457)
(4.4,0.344363887)
(4.5,0.355361181)
(4.6,0.366343158)
(4.7,0.377297748)
(4.8,0.388211293)
(4.9,0.399071473)
(5.0,0.409865635)
};

\addlegendentry{$[30,60]$\,m/s}

\node[
    anchor=south east
] at (axis cs:5,0.01)
{mean $J^*_\ell$};

\addplot[
    black,
    dashed,
    domain=-5:5,
    samples=2
] {0.010858493001};
\end{axis}

\end{tikzpicture}
\caption{\textbf{Mean weighted spectral consistency score} (\ref{eq:weighted_nll}) as a function of the \textbf{velocity deviation} $\overline{v}_\ell-\tilde{v}_\ell$ for different ground-truth velocity intervals (color code) in $\mathcal{D}^{\mathrm{test}}$.}
\label{fig:nll_velocity_deviation}
\end{figure}

%% file: figures/results_auroc.tex
\begin{table*}[t]
    \centering
    \caption{\textbf{AUROC for detecting anomalies in Doppler spectra}
    using conventional spectral baselines (left table segment) and our proposed anomaly scores (right table segment) on $\mathcal{D}^{\mathrm{test}}$.
    Higher values indicate better discrimination. Best metrics are bold, second best are underlined.}
    \label{tab:perturbation_auroc}
    \setlength{\tabcolsep}{4pt}
    
\begin{tabular}{l|l|rrr|rrr}
\toprule
Anomaly & Strength &
\makecell{\\\texttt{SE} \cite{shannon1948mathematical}} &
\makecell{Baselines\\\texttt{-PNR} \cite{meyer2004peaknoise}} &
\makecell{\\\texttt{VCSE}} &
\makecell{\\SAS (\ref{eq:spectral_anomaly})} &
\makecell{Ours\\CAS (\ref{eq:conditional_anomaly})} &
\makecell{\\AVD (\ref{eq:velocity_disagreement})} \\
\midrule

\multirow{3}{*}{\textsf{Shift}} & \phantom{1}4 bins &
0.500 & 0.498 & 0.563 &
\underline{0.752} & \textbf{0.759} & 0.686 \\

& \phantom{1}8 bins &
0.500 & 0.499 & 0.611 &
\textbf{0.873} & \underline{0.871} & 0.639 \\

& 16 bins &
0.500 & 0.505 & 0.692 &
\underline{0.930} & \textbf{0.934} & 0.809 \\

\midrule

\multirow{3}{*}{\textsf{Extra Peak}} & 0.1 &
0.629 & 0.504 & 0.861 &
\underline{0.965} & \textbf{0.969} & 0.897 \\

& 0.3 &
0.703 & 0.503 & 0.975 &
\underline{0.979} & \textbf{0.986} & 0.959 \\

& 0.6 &
0.715 & 0.493 & \underline{0.991} &
0.986 & \textbf{0.993} & 0.977 \\

\midrule

\multirow{3}{*}{\textsf{Local}} & 0.1 &
0.632 & 0.503 & 0.868 &
\underline{0.966} & \textbf{0.971} & 0.898 \\

& 0.3 &
0.710 & 0.501 & 0.978 &
\underline{0.979} & \textbf{0.988} & 0.966 \\

& 0.6 &
0.728 & 0.474 & \underline{0.992} &
0.986 & \textbf{0.994} & 0.985 \\
\midrule

\multirow{3}{*}{\textsf{Attenuation}} & 0.25 &
\underline{0.989} & \textbf{1.000} & 0.879 &
\textbf{1.000} & \textbf{1.000} & 0.970 \\

& 0.50 &
\underline{0.994} & \textbf{1.000} & 0.981 &
\textbf{1.000} & \textbf{1.000} & 0.992 \\

& 0.90 &
0.994 & \textbf{1.000} & 0.993 &
\textbf{1.000} & \textbf{1.000} & \underline{0.998} \\
\midrule

\multirow{7}{*}{\textsf{Noise}} & \hspace{-0.8mm}$-5$ dB &
\textbf{1.000} & \textbf{1.000} & \underline{0.993} &
\textbf{1.000} & \textbf{1.000} & \textbf{1.000} \\

& \phantom{1}0 dB &
\textbf{1.000} & \textbf{1.000} & 0.985 &
\textbf{1.000} & \textbf{1.000} & \underline{0.999} \\

& \phantom{1}5 dB &
\underline{0.996} & \textbf{1.000} & 0.884 &
\textbf{1.000} & \textbf{1.000} & 0.994 \\

& 10 dB &
0.838 & \underline{0.998} & 0.689 &
\textbf{1.000} & \textbf{1.000} & 0.948 \\

& 15 dB &
0.641 & \underline{0.993} & 0.569 &
\textbf{0.998} & \textbf{0.998} & 0.765 \\

& 20 dB &
0.549 & 0.947 & 0.522 &
\textbf{0.985} & \underline{0.984} & 0.585 \\

& 30 dB &
0.505 & 0.625 & 0.502 &
\textbf{0.646} & \underline{0.644} & 0.502 \\

\bottomrule
\end{tabular}

\end{table*}

%% file: figures/results_scatter.tex
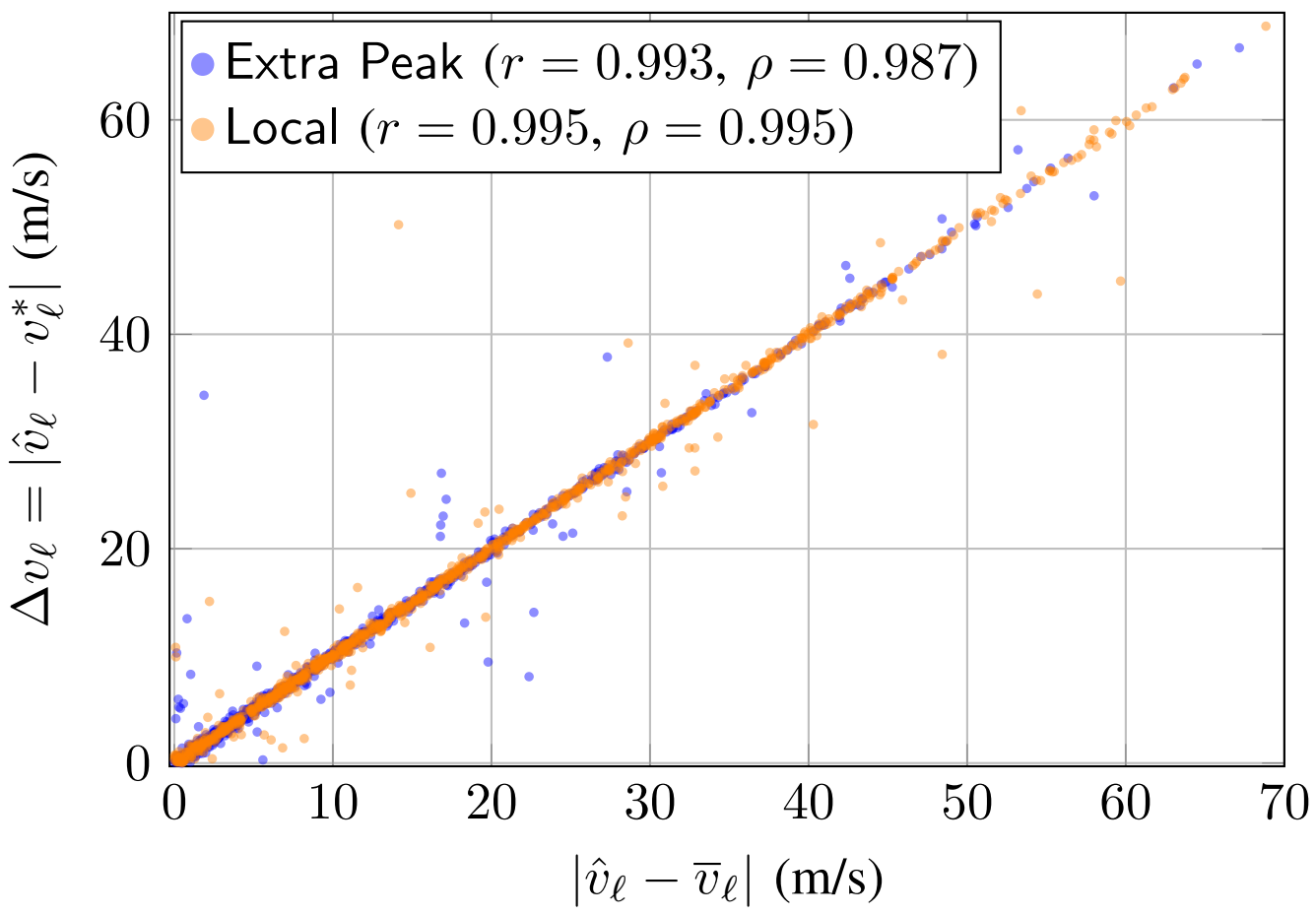
\begin{figure}[t]
\centering
\begin{tikzpicture}
\begin{axis}[
 width=0.5\textwidth,height=0.75\columnwidth,
 xlabel={$|\hat v_\ell-\overline{v}_\ell|$ (m/s)},
 ylabel={$\Delta v_\ell=|\hat v_\ell-v_\ell^*|$ (m/s)},
 grid=major,
 xmin=-0.3,xmax=70,
 ymin=-0.3, ymax=70,
 legend image post style={mark size=2pt},
 legend pos=north west,
    legend cell align=left,
    legend style={
    inner sep=2pt,
    at={(0.01,0.99)},
        fill=white,
        draw=black
    }
]
\addplot[only marks,mark=*,mark size=0.7pt,blue,opacity=0.45] coordinates {
 (4.03303,4.08355)
 (14.83511,14.67524)
 (3.05601,2.83893)
 (4.27416,4.23739)
 (4.29580,4.06752)
 (7.99841,8.12885)
 (18.20341,18.16089)
 (20.93810,21.11547)
 (21.59681,21.59903)
 (13.10248,13.26430)
 (1.45363,2.12508)
 (20.42368,20.56500)
 (8.15969,8.03632)
 (30.08607,30.02216)
 (0.94790,1.59493)
 (40.63271,40.75212)
 (2.00828,2.65000)
 (0.24050,0.18052)
 (35.36136,34.74144)
 (20.19111,20.16898)
 (31.02347,30.88387)
 (27.93821,28.15192)
 (17.01323,16.97238)
 (3.74279,3.45277)
 (16.77397,15.76564)
 (12.63993,12.49577)
 (4.06326,3.98702)
 (5.22344,5.24015)
 (7.20570,7.25613)
 (11.52417,11.53714)
 (32.76969,32.50652)
 (36.81687,36.66566)
 (1.46482,1.35535)
 (2.25052,2.33174)
 (31.89217,31.45883)
 (0.59271,0.79754)
 (26.96729,27.25673)
 (5.92417,5.80860)
 (29.56879,29.38781)
 (7.78856,7.70966)
 (10.50243,10.68077)
 (2.07359,1.91679)
 (7.33745,7.38535)
 (18.26828,17.96042)
 (19.55622,19.39793)
 (29.61396,29.35699)
 (23.26280,23.32771)
 (40.63715,40.86750)
 (20.01785,20.55237)
 (7.88982,7.78277)
 (5.65880,5.48390)
 (17.31934,17.39488)
 (30.42675,30.17619)
 (25.24903,25.16129)
 (6.37143,6.25564)
 (5.74243,5.77302)
 (2.92243,1.83142)
 (19.53052,19.71832)
 (6.78477,6.41035)
 (2.02607,2.32471)
 (37.26683,37.26513)
 (7.13531,6.96106)
 (12.37850,12.35501)
 (16.38522,16.43456)
 (6.83378,6.62624)
 (7.05570,7.09010)
 (0.70280,0.88351)
 (72.52106,69.44934)
 (29.59363,29.38324)
 (15.96724,15.80379)
 (0.78315,0.57082)
 (9.18224,9.05485)
 (3.20118,2.91055)
 (22.63090,23.19640)
 (43.00929,42.77195)
 (1.21334,1.10825)
 (1.18239,1.51926)
 (32.92810,32.87851)
 (1.25559,1.28527)
 (5.73157,5.89896)
 (1.80217,1.82052)
 (14.24424,14.48494)
 (25.82796,25.77450)
 (6.22869,6.06833)
 (2.31084,2.38815)
 (2.21874,1.50280)
 (6.33965,6.29854)
 (3.06734,3.49751)
 (8.30946,8.38412)
 (5.66230,5.61941)
 (1.16909,1.45992)
 (17.04429,17.24866)
 (14.93513,14.55445)
 (13.55687,13.58298)
 (10.44668,11.06763)
 (5.20719,9.03939)
 (6.08897,6.11990)
 (5.37441,6.15390)
 (31.33315,31.18481)
 (3.29166,3.06949)
 (34.28370,34.10106)
 (5.60349,6.29672)
 (14.31883,14.05935)
 (9.70785,9.66336)
 (11.31923,11.16248)
 (1.68012,1.50086)
 (19.94385,20.75915)
 (1.84414,1.31367)
 (4.12690,3.80394)
 (9.39902,9.34084)
 (4.43642,4.50533)
 (8.30870,8.40139)
 (7.29825,6.85136)
 (3.38571,2.93263)
 (12.82222,12.73119)
 (14.95951,14.87984)
 (10.02593,9.91076)
 (23.49976,23.20621)
 (16.68143,16.96339)
 (50.54553,50.13480)
 (12.71281,12.72615)
 (10.15775,10.13998)
 (1.51536,1.76205)
 (70.19033,70.03709)
 (21.32685,21.90625)
 (4.02278,3.20869)
 (0.27637,0.32027)
 (46.32579,46.09549)
 (5.25280,5.33891)
 (4.49868,4.53643)
 (22.95024,23.15601)
 (5.17480,5.27469)
 (8.91560,8.66343)
 (37.21159,36.99767)
 (2.07729,1.88485)
 (19.55618,19.34521)
 (0.41670,0.61151)
 (0.10488,0.19658)
 (30.70773,30.77457)
 (3.98156,4.00868)
 (5.76269,5.51852)
 (1.40185,1.33238)
 (47.64287,47.42910)
 (26.79437,27.45872)
 (50.46541,50.28975)
 (2.29670,2.30444)
 (12.58152,12.69017)
 (17.14575,24.60823)
 (1.57976,0.86974)
 (14.74957,14.09492)
 (35.93913,35.78722)
 (4.28594,4.20072)
 (2.81879,2.85213)
 (2.82581,2.98475)
 (4.08686,4.08622)
 (9.97612,10.15187)
 (4.53430,4.66870)
 (11.70208,11.82301)
 (5.37538,5.38242)
 (1.89195,1.84792)
 (13.10582,12.86589)
 (1.22067,1.72875)
 (29.00097,28.94483)
 (4.08344,4.25236)
 (0.99237,0.72885)
 (26.53972,26.53725)
 (5.73716,5.99817)
 (4.57412,4.95390)
 (6.44168,6.21241)
 (4.07962,4.15463)
 (13.69412,13.46870)
 (13.37816,13.46161)
 (18.79113,18.45841)
 (10.58788,10.47841)
 (2.86651,3.05326)
 (12.80457,13.30158)
 (28.50391,28.52953)
 (13.65312,14.03601)
 (0.33917,0.18967)
 (0.36085,0.39380)
 (3.31855,3.38625)
 (11.42119,11.02831)
 (6.23378,6.14677)
 (9.14325,9.02164)
 (21.79815,21.58680)
 (4.73602,4.71018)
 (8.80916,8.95722)
 (6.52190,6.26741)
 (1.85516,2.01632)
 (22.67226,14.05772)
 (3.11079,3.46061)
 (11.15739,11.06081)
 (1.01564,0.91889)
 (6.98274,6.50071)
 (19.69533,16.87230)
 (8.80464,8.94696)
 (0.71492,0.61478)
 (27.10721,27.56578)
 (0.84669,0.53140)
 (26.54558,27.06947)
 (21.25825,21.19605)
 (21.64700,21.54513)
 (6.54921,6.12511)
 (28.37859,28.16684)
 (15.92339,16.17017)
 (24.51252,21.16014)
 (27.26787,27.54469)
 (16.75658,16.90871)
 (43.38212,43.33501)
 (2.59856,2.54815)
 (35.78169,35.62089)
 (10.76738,10.90901)
 (13.31211,13.24244)
 (3.04807,3.01380)
 (1.26278,1.09351)
 (9.48125,9.48856)
 (0.79179,0.74677)
 (23.87147,24.17199)
 (2.40939,2.34501)
 (36.47421,36.31226)
 (2.71743,2.29382)
 (10.83901,10.93979)
 (8.19763,8.19080)
 (9.19697,9.48733)
 (4.65144,4.65541)
 (29.14688,29.11677)
 (18.10193,18.04993)
 (6.67460,6.62243)
 (6.47425,6.30070)
 (5.01026,5.43727)
 (54.19544,54.23498)
 (8.31507,8.26631)
 (14.60693,14.83434)
 (7.12312,7.05334)
 (0.28705,5.21945)
 (14.89504,14.79935)
 (6.81796,6.67389)
 (24.29989,24.16538)
 (5.14254,4.88298)
 (53.76073,53.60080)
 (4.16420,4.45606)
 (2.53315,2.69638)
 (0.91060,0.79449)
 (8.36576,7.34220)
 (0.74149,0.76160)
 (10.58930,10.62218)
 (10.66191,11.11625)
 (4.50662,4.53729)
 (15.01155,14.84192)
 (33.98076,34.02339)
 (7.72830,7.77857)
 (20.88073,21.64143)
 (7.86314,7.88990)
 (10.16937,10.58824)
 (18.60931,18.29629)
 (6.33615,6.08626)
 (9.19112,9.13990)
 (4.57570,4.48336)
 (6.70589,6.85612)
 (8.20796,7.22404)
 (26.73210,27.18150)
 (38.68658,38.53623)
 (0.91182,1.75326)
 (0.48893,0.16826)
 (10.99191,10.78874)
 (1.31655,0.96938)
 (19.21802,19.06243)
 (1.97003,1.86453)
 (8.36870,8.91496)
 (4.86619,5.17699)
 (0.41669,0.41872)
 (12.34969,11.10326)
 (12.49668,13.70543)
 (7.76315,7.71872)
 (0.82009,0.78068)
 (3.50698,3.59563)
 (8.51456,8.82095)
 (10.32382,9.33574)
 (2.58933,3.00600)
 (38.24998,38.14492)
 (13.29567,13.35581)
 (19.88936,19.67541)
 (7.94522,7.64631)
 (0.09641,0.26063)
 (8.85492,9.58801)
 (1.87181,34.30947)
 (32.58013,32.80146)
 (0.21971,0.08742)
 (3.08485,3.27416)
 (44.08760,43.89476)
 (8.45527,8.65767)
 (6.10472,5.89913)
 (6.46336,6.85120)
 (6.78209,6.79686)
 (3.49754,3.58168)
 (8.06990,8.33979)
 (4.59446,4.81246)
 (9.26112,9.02950)
 (5.87160,6.51095)
 (1.13844,1.29068)
 (1.93676,2.21579)
 (3.20064,2.90700)
 (5.37783,5.57048)
 (28.69624,28.29592)
 (18.18264,18.11012)
 (2.07574,1.84166)
 (30.89954,31.09327)
 (6.61846,6.97344)
 (21.16996,21.57946)
 (42.05670,42.43662)
 (3.85651,3.22417)
 (0.33764,0.14592)
 (3.85643,3.76005)
 (2.45136,2.24474)
 (11.83899,11.85433)
 (7.64440,7.57841)
 (7.96346,8.44218)
 (9.80490,10.18088)
 (19.36378,19.38100)
 (43.24368,42.68457)
 (8.27519,8.21712)
 (0.03861,0.23714)
 (9.97674,10.04166)
 (6.03102,5.97116)
 (3.37357,2.77746)
 (16.84299,27.03912)
 (11.68023,11.62990)
 (3.19614,3.90565)
 (5.15145,5.27403)
 (1.36612,1.24888)
 (9.52938,9.52189)
 (9.91283,9.66325)
 (18.96770,19.12677)
 (3.43700,3.08395)
 (8.08294,8.10468)
 (11.59177,11.37076)
 (4.68623,4.80911)
 (25.07632,25.15992)
 (30.21870,30.26044)
 (11.83740,11.20454)
 (21.78571,21.63346)
 (9.30780,9.27614)
 (3.16110,3.07885)
 (3.69796,3.50111)
 (12.42059,12.49387)
 (2.78966,2.35020)
 (9.41602,9.51582)
 (25.20583,25.20219)
 (1.78464,0.91446)
 (5.63428,5.49226)
 (10.10468,10.48778)
 (4.35250,5.00472)
 (33.54199,34.45683)
 (23.42643,23.61293)
 (0.81154,0.80230)
 (3.75967,3.80701)
 (3.20034,3.48882)
 (3.20580,3.46421)
 (7.84431,7.73801)
 (27.26469,27.28062)
 (23.12748,23.15355)
 (23.03596,23.34916)
 (5.05022,5.15650)
 (6.00101,6.05828)
 (5.22670,2.90463)
 (20.98194,20.99871)
 (13.95512,14.00198)
 (2.95396,2.75314)
 (15.40550,15.59025)
 (1.60334,1.54903)
 (8.79520,8.74230)
 (4.59317,4.61200)
 (1.16404,1.60966)
 (6.11691,6.07173)
 (4.35983,4.11069)
 (8.00857,8.13002)
 (20.41567,20.51507)
 (47.07406,47.24409)
 (19.25796,19.44027)
 (2.78738,2.63913)
 (9.07777,8.89001)
 (2.48064,3.15036)
 (42.68408,42.56088)
 (27.76798,27.90487)
 (42.31413,42.34820)
 (43.39270,43.26442)
 (6.33121,5.88164)
 (16.95720,23.03753)
 (19.62548,19.14515)
 (1.66326,2.18077)
 (0.62950,0.82091)
 (7.76830,7.79039)
 (0.25538,5.95147)
 (0.11776,0.60408)
 (2.79422,2.68879)
 (32.04079,32.04990)
 (2.48812,2.64847)
 (3.19641,2.55064)
 (4.92155,4.73711)
 (0.19057,0.16680)
 (4.06411,4.00550)
 (2.29476,2.17707)
 (28.02808,27.35319)
 (1.78320,1.55421)
 (10.71455,10.81308)
 (4.56242,4.11996)
 (2.22882,2.56393)
 (53.20221,57.21216)
 (7.87377,7.94368)
 (4.99787,4.97820)
 (1.09019,1.02101)
 (7.61917,7.84646)
 (10.16206,10.08271)
 (0.09964,4.14157)
 (32.01595,31.91247)
 (13.30289,13.57373)
 (44.87851,44.85801)
 (5.89607,5.94753)
 (20.19356,20.89783)
 (63.01324,62.98718)
 (15.94380,16.23592)
 (16.39287,15.86887)
 (18.97819,18.69791)
 (13.07766,13.27877)
 (3.04045,2.75318)
 (18.54022,18.31722)
 (1.21314,0.59478)
 (5.88353,5.77290)
 (3.69843,3.63535)
 (18.07734,17.46599)
 (3.63426,3.33270)
 (16.40348,16.05903)
 (28.53507,25.31280)
 (4.81380,4.70226)
 (0.47813,0.50052)
 (4.68172,3.83978)
 (45.27500,44.40226)
 (3.96771,3.63276)
 (3.33527,2.96123)
 (1.08060,1.20811)
 (7.05439,7.21615)
 (26.31956,26.41031)
 (9.52963,9.97429)
 (5.83432,5.64420)
 (4.94770,5.24929)
 (1.60365,1.46830)
 (5.38006,5.43529)
 (3.32516,3.60500)
 (2.69662,2.21135)
 (17.95925,17.72770)
 (12.90133,12.69034)
 (4.60447,4.90661)
 (9.73385,9.86676)
 (8.27612,7.50501)
 (18.56165,18.68223)
 (43.79323,43.98381)
 (15.15887,15.31623)
 (1.03413,8.27384)
 (28.08511,28.09955)
 (3.30173,3.34277)
 (15.28406,15.14410)
 (12.90157,12.73308)
 (4.61391,4.62885)
 (0.36518,0.70020)
 (18.30567,13.06383)
 (4.27082,4.53129)
 (9.81098,10.26393)
 (0.62650,0.42751)
 (35.81023,35.96507)
 (0.44958,0.59428)
 (5.76930,5.90844)
 (24.76788,24.88450)
 (28.51581,28.05721)
 (1.01381,0.10145)
 (2.62770,2.48656)
 (0.55213,0.38489)
 (2.55708,2.07320)
 (8.78698,8.79311)
 (29.69112,29.73350)
 (3.58258,3.84821)
 (24.90267,24.90294)
 (17.79611,17.55525)
 (36.62761,36.32067)
 (1.04493,1.06956)
 (11.75559,11.56661)
 (4.16721,3.92697)
 (3.40270,2.98036)
 (2.79826,2.32725)
 (5.19000,4.80955)
 (5.13701,5.38145)
 (4.85016,4.68427)
 (2.54853,1.90487)
 (15.53207,15.37199)
 (17.25142,16.54504)
 (30.61291,30.52350)
 (25.72964,25.57700)
 (1.97473,0.96826)
 (5.00421,4.84283)
 (0.24298,0.68015)
 (30.59492,29.53069)
 (1.84708,1.52459)
 (6.10700,6.36444)
 (38.93272,38.97947)
 (17.26645,17.19738)
 (13.55453,13.94510)
 (13.12094,13.37933)
 (3.78306,3.88349)
 (2.67392,3.07506)
 (0.80557,13.46778)
 (41.98966,41.23600)
 (7.36110,7.41700)
 (2.45610,2.48149)
 (1.03163,1.01687)
 (0.79435,0.81002)
 (11.68682,11.73309)
 (3.91277,3.64695)
 (8.60756,8.71558)
 (3.96942,3.53483)
 (0.25172,0.42416)
 (17.42133,16.93390)
 (16.75865,16.69493)
 (44.80545,44.86357)
 (11.94091,12.21901)
 (23.96405,24.18431)
 (7.45679,7.35325)
 (5.15063,5.10629)
 (33.87408,33.35913)
 (6.88051,7.19756)
 (0.55822,0.27133)
 (3.38668,2.79929)
 (13.07009,13.05628)
 (1.43396,0.98550)
 (3.94352,3.50087)
 (1.74298,1.78890)
 (1.56669,1.99699)
 (0.61897,0.60773)
 (16.77960,21.14635)
 (2.10410,2.42434)
 (1.94646,2.09655)
 (34.65343,34.53135)
 (22.48645,22.26659)
 (0.71032,0.69236)
 (39.54451,39.10995)
 (2.94912,2.95301)
 (19.92154,19.42148)
 (0.35494,0.47942)
 (0.14951,10.27452)
 (1.93898,2.23886)
 (11.00882,11.39159)
 (23.13137,23.09404)
 (17.81219,17.69284)
 (11.33109,11.72636)
 (5.34220,5.76192)
 (5.86135,5.71700)
 (56.35853,56.42733)
 (2.11745,1.80463)
 (13.97877,13.96618)
 (11.11655,11.22506)
 (7.62897,7.17316)
 (6.37422,6.60593)
 (0.00942,0.04227)
 (3.61615,3.68925)
 (19.17306,19.06631)
 (34.58934,34.42043)
 (0.61955,0.54556)
 (13.64290,14.01305)
 (5.11998,5.02708)
 (22.62754,21.72791)
 (4.78234,4.62938)
 (5.58773,6.10392)
 (4.70940,4.19238)
 (2.92544,2.74756)
 (5.72948,6.22746)
 (6.80369,6.73552)
 (65.99673,76.56734)
 (9.81671,6.59832)
 (40.29083,40.26595)
 (5.79171,5.88900)
 (9.55262,9.86749)
 (2.52736,2.76112)
 (0.34860,0.27096)
 (3.81356,3.50506)
 (5.94780,6.03305)
 (1.77402,1.98642)
 (5.30646,5.53599)
 (10.69563,10.73257)
 (16.36097,16.90728)
 (2.22210,2.41334)
 (3.67883,4.84151)
 (18.79284,18.89627)
 (11.30596,11.10467)
 (3.34889,3.18538)
 (29.31480,29.54771)
 (10.89239,11.34924)
 (31.37234,31.59797)
 (5.52329,5.31305)
 (0.08797,0.19960)
 (76.41222,76.23967)
 (20.30785,19.87979)
 (7.75422,7.58398)
 (22.18097,22.41223)
 (10.52620,10.44544)
 (1.86368,2.13672)
 (4.28209,4.36634)
 (12.78417,13.14836)
 (29.50433,29.42966)
 (1.52866,3.39490)
 (0.65402,0.42382)
 (13.38436,13.32535)
 (21.67751,21.75754)
 (1.76082,2.33456)
 (8.41793,8.20744)
 (3.80488,3.83209)
 (0.47605,0.22361)
 (4.84251,4.90580)
 (13.39878,12.77275)
 (10.04469,10.28244)
 (1.03078,0.96830)
 (12.19448,12.30088)
 (14.32550,14.14091)
 (7.71730,7.57546)
 (12.99165,12.93990)
 (6.94353,7.07535)
 (19.44284,19.25667)
 (0.27358,0.57960)
 (64.49652,65.21095)
 (17.60250,17.55479)
 (8.45000,8.42308)
 (8.96594,8.74139)
 (5.54046,5.36117)
 (4.60390,5.04585)
 (20.29840,19.74267)
 (41.97720,42.05613)
 (17.64382,17.30697)
 (2.87166,2.48485)
 (16.16910,16.90015)
 (17.35106,17.01534)
 (2.91216,2.78994)
 (17.83005,17.65731)
 (10.06579,10.09979)
 (3.32481,3.31749)
 (38.07883,38.25700)
 (12.38486,11.83841)
 (3.19065,3.40600)
 (11.35281,11.23799)
 (7.63159,7.94153)
 (18.62857,18.84417)
 (32.27187,32.27368)
 (23.85532,22.31399)
 (16.39884,16.04358)
 (9.24708,5.94693)
 (6.06070,6.04794)
 (19.64284,19.52492)
 (19.77972,19.38243)
 (26.34410,26.52581)
 (4.68360,4.61883)
 (34.10164,33.46861)
 (27.96815,28.75704)
 (3.84610,3.91619)
 (10.77114,11.42259)
 (0.57981,5.54728)
 (1.44018,1.40337)
 (0.36814,0.15783)
 (2.34300,2.02577)
 (1.62637,1.53893)
 (23.57606,23.74932)
 (18.52210,18.51619)
 (2.44566,2.82211)
 (1.76783,2.02774)
 (26.62319,26.50982)
 (12.07607,12.02220)
 (20.36852,19.61039)
 (11.99341,11.96674)
 (31.72003,31.28416)
 (31.81200,31.82534)
 (49.00934,49.52605)
 (26.44746,26.35460)
 (25.98422,26.44204)
 (5.80119,5.60408)
 (10.35824,9.91967)
 (1.99191,1.99206)
 (0.41678,0.43134)
 (3.61567,3.41565)
 (1.53493,1.27647)
 (10.34225,10.54775)
 (3.32240,3.84497)
 (17.16843,16.62708)
 (8.38174,8.33390)
 (7.20481,7.19611)
 (10.85956,11.29218)
 (6.62649,6.54750)
 (3.42821,3.28057)
 (3.78382,3.56314)
 (3.29845,3.25880)
 (5.18439,5.20508)
 (0.19609,0.38741)
 (1.20205,1.15276)
 (15.49245,15.44651)
 (55.26310,55.49963)
 (16.02985,15.94863)
 (5.19556,5.71704)
 (7.18060,6.90364)
 (3.49744,4.28694)
 (2.75849,2.89450)
 (13.31388,13.10041)
 (9.45735,9.51323)
 (6.25653,5.97014)
 (27.83221,27.38408)
 (33.43638,33.81462)
 (18.66556,18.43231)
 (11.92688,12.24299)
 (2.79820,3.16216)
 (5.74408,5.74846)
 (18.30839,18.71722)
 (1.94100,2.01561)
 (36.41550,32.68675)
 (3.76111,3.72013)
 (0.22539,0.48098)
 (3.87834,4.00693)
 (5.43630,5.31089)
 (14.91301,14.56205)
 (34.83384,34.54611)
 (9.14028,9.28974)
 (5.10079,4.97969)
 (35.16325,35.00465)
 (30.27108,30.28765)
 (0.55418,0.23422)
 (3.22851,3.06498)
 (4.75975,4.61498)
 (18.22396,18.23001)
 (0.52953,0.45920)
 (5.77458,5.86093)
 (26.55481,26.97353)
 (15.49452,15.45037)
 (8.39807,9.00325)
 (10.71595,10.88353)
 (20.76127,21.06710)
 (12.51158,12.47059)
 (5.17085,5.34873)
 (52.57983,51.82990)
 (3.94103,4.26072)
 (11.00336,11.10477)
 (27.30534,37.87637)
 (67.16272,66.72192)
 (1.64305,1.61876)
 (15.70249,15.11699)
 (22.60116,22.55781)
 (6.18990,5.95242)
 (1.56512,1.68255)
 (15.49789,15.26043)
 (8.68852,8.98512)
 (1.75681,1.48268)
 (30.71405,27.08327)
 (18.62874,18.80546)
 (8.75798,8.60533)
 (15.59662,15.60582)
 (8.85942,8.82931)
 (20.19836,20.62731)
 (0.38849,5.10913)
 (11.31331,11.58298)
 (15.48262,16.01809)
 (11.60217,11.47208)
 (1.27539,0.92883)
 (29.00085,28.84232)
 (10.23122,10.10232)
 (25.76031,26.27021)
 (8.92397,8.98988)
 (0.71830,0.72449)
 (3.11060,2.83828)
 (5.30039,5.34192)
 (1.70089,2.11144)
 (19.26126,19.08642)
 (40.96114,40.85387)
 (14.87057,15.02684)
 (3.43650,3.46010)
 (1.24420,1.65002)
 (42.60003,45.22482)
 (10.50601,10.87382)
 (8.10419,8.27732)
 (10.25330,10.47343)
 (1.53826,1.45444)
 (70.18822,70.23383)
 (26.84654,27.02208)
 (2.41598,1.67086)
 (15.99783,15.87386)
 (42.34500,46.41288)
 (12.07256,12.23582)
 (1.63886,1.55734)
 (50.63972,50.93185)
 (25.12295,21.45039)
 (2.24207,2.07856)
 (11.88543,12.57835)
 (2.59307,2.84354)
 (12.59189,12.98658)
 (8.93433,8.86536)
 (6.83287,6.71025)
 (0.96174,0.73513)
 (25.57894,25.85239)
 (20.47859,20.63377)
 (3.70117,3.72493)
 (8.88020,10.24941)
 (8.82245,8.09447)
 (2.91607,3.03138)
 (12.40794,12.60427)
 (6.12497,6.55820)
 (4.20248,3.89004)
 (5.58503,0.29609)
 (17.96916,17.94668)
 (0.42754,0.05202)
 (1.28984,0.54337)
 (11.53778,11.51970)
 (17.63262,17.50008)
 (9.77089,9.99671)
 (7.67217,7.70295)
 (5.58992,5.80870)
 (1.25105,1.37783)
 (2.47604,2.23750)
 (32.37628,32.24296)
 (24.79914,24.99147)
 (3.31222,3.44851)
 (37.26237,36.97673)
 (22.63343,22.65479)
 (21.39482,21.36668)
 (7.82657,7.90076)
 (31.42922,31.19970)
 (13.79055,14.02920)
 (3.61035,3.22663)
 (1.87112,1.84163)
 (13.19477,13.65039)
 (7.00186,7.10793)
 (10.91448,10.93046)
 (12.76114,13.08111)
 (12.76639,12.48702)
 (3.70640,4.49074)
 (0.96490,1.37546)
 (5.05036,4.46653)
 (3.70850,3.40408)
 (1.03433,1.23151)
 (18.34983,18.52048)
 (13.03345,13.49750)
 (6.35949,6.26176)
 (12.46476,12.36472)
 (0.89769,1.36095)
 (3.49160,3.49318)
 (2.98837,3.05099)
 (6.05161,5.92270)
 (6.36570,6.43158)
 (1.17694,1.31487)
 (20.80464,20.39301)
 (23.14855,22.96651)
 (3.96436,3.96175)
 (6.64911,6.58984)
 (6.83651,7.04017)
 (27.84172,28.17284)
 (1.91124,1.59460)
 (4.49161,4.64859)
 (2.47310,2.71206)
 (38.21598,38.02063)
 (7.56206,7.42827)
 (6.28330,6.27440)
 (0.49652,1.39887)
 (1.63206,1.52788)
 (1.35475,1.69892)
 (15.37618,15.04690)
 (4.51518,4.45894)
 (0.76220,0.47823)
 (5.89442,6.21422)
 (19.17952,19.68916)
 (13.50136,13.50601)
 (4.08949,4.32975)
 (1.70045,2.00705)
 (7.95715,7.89075)
 (27.12710,27.10146)
 (20.74968,20.77424)
 (8.23071,8.23792)
 (4.07998,3.91708)
 (7.03114,7.01961)
 (0.10476,0.12035)
 (30.16822,30.16980)
 (1.33668,1.39410)
 (3.89144,3.87514)
 (10.90314,10.91402)
 (2.11722,1.47799)
 (25.68472,25.70492)
 (5.56872,5.51848)
 (10.53354,10.43629)
 (41.92272,41.56118)
 (42.58018,42.87065)
 (26.93585,26.79035)
 (0.98394,0.56460)
 (29.41213,29.29572)
 (1.49222,1.46564)
 (4.57381,4.13288)
 (3.81456,3.89103)
 (0.48680,0.00054)
 (2.40715,2.71788)
 (25.42134,25.62430)
 (16.90702,16.85598)
 (16.71806,17.18044)
 (26.62580,26.39552)
 (23.83561,24.52792)
 (27.06595,26.97260)
 (10.01261,9.83696)
 (16.58770,16.44951)
 (24.18939,24.39169)
 (0.37421,0.64583)
 (5.93290,5.65101)
 (12.31909,12.35221)
 (5.61490,5.58813)
 (11.45219,11.39644)
 (16.10966,16.19671)
 (9.12740,9.05212)
 (2.67155,2.94268)
 (16.80013,22.21024)
 (14.63441,15.13877)
 (6.80320,6.86906)
 (6.43123,6.38567)
 (14.06162,13.93602)
 (16.07355,15.80873)
 (26.77845,26.66100)
 (7.18639,8.23742)
 (0.39045,0.03019)
 (0.95065,0.69262)
 (10.96544,10.35961)
 (58.00302,52.91719)
 (19.78858,9.42940)
 (11.82553,12.18118)
 (0.72660,0.69935)
 (3.90914,3.84889)
 (15.34741,15.20708)
 (13.81849,13.25360)
 (9.31967,9.43259)
 (15.81300,16.08246)
 (4.66469,4.86979)
 (41.06364,41.13117)
 (48.62347,48.63673)
 (4.00825,4.03909)
 (6.54746,6.46592)
 (6.03995,6.19294)
 (22.37030,8.06226)
 (13.48833,13.02642)
 (4.80523,4.98893)
 (2.51114,2.75237)
 (5.60115,5.86064)
 (4.87065,5.13883)
 (12.70436,12.84496)
 (27.52622,26.91319)
 (0.93957,0.90391)
 (1.10530,0.74556)
 (7.13546,7.16243)
 (28.31634,28.70290)
 (24.49384,24.94951)
 (7.00788,6.83843)
 (8.17310,8.14375)
 (19.35618,19.05981)
 (3.30250,3.03613)
 (13.79337,14.00730)
 (3.70296,3.48005)
 (48.40263,47.97181)
 (48.41112,50.77278)
 (7.51259,8.00407)
 (31.22438,31.08114)
 (2.67455,2.19116)
 (31.45571,31.25880)
 (5.68122,5.70730)
 (37.25106,37.00529)
 (14.83395,14.45503)
 (40.82034,40.86784)
 (0.10463,0.55083)
 (12.89169,14.29823)
 (2.50733,2.84516)
 (10.76212,10.51536)
 (6.48836,5.16056)
 (31.39856,31.09653)
 (24.12516,23.97568)
 (7.63644,7.50715)
 (0.36446,0.05292)
 (5.67101,5.90525)
 (6.10386,5.58665)
 (17.54241,17.77205)
 (39.17650,39.42488)
 (13.62613,13.68950)
 (1.70424,1.59404)
 (5.13163,4.87837)
 (10.83919,10.77089)
 (2.52590,2.01014)
 (5.70237,4.72071)
 (2.69541,2.59063)
 (5.27553,5.25182)
 (6.14384,5.96069)
 (44.57815,44.63813)
 (29.92485,30.09341)
 (0.24408,0.64241)
 (1.35474,0.64005)
 (8.24200,7.89301)
 (7.73022,7.77783)
 (10.19347,10.05635)
 (2.53415,2.02332)
 (6.08829,6.15411)
 (7.79726,8.08414)
};
\addlegendentry{\textsf{Extra Peak} ($r=0.993$, $\rho=0.987$)}
\addplot[only marks,mark=*,mark size=0.7pt,orange,opacity=0.45] coordinates {
 (3.67025,3.51973)
 (43.03201,42.87215)
 (33.78385,33.80093)
 (5.88532,5.84855)
 (7.28620,7.05792)
 (29.42916,29.45961)
 (3.47417,3.43165)
 (52.39645,52.57382)
 (45.30925,45.31148)
 (25.52860,25.69042)
 (0.27982,0.49163)
 (37.66879,37.81010)
 (32.75914,32.63577)
 (1.81081,1.84690)
 (24.70315,24.15612)
 (10.68044,10.69986)
 (9.14181,9.78353)
 (32.33159,32.27161)
 (11.17584,8.65592)
 (38.62158,38.59945)
 (12.64979,12.51019)
 (22.23394,22.44765)
 (17.94832,18.00747)
 (13.20299,12.91297)
 (25.43519,24.72686)
 (42.54463,42.40047)
 (7.70381,7.78006)
 (10.55371,10.57042)
 (5.55590,5.60632)
 (4.90804,4.79508)
 (33.15494,33.51812)
 (32.81613,32.66492)
 (3.28169,3.07223)
 (18.24249,18.22371)
 (0.49931,0.03403)
 (35.82877,36.13361)
 (16.92332,17.11275)
 (0.90714,0.79157)
 (26.76572,26.64670)
 (18.82620,18.74730)
 (19.16225,19.34060)
 (4.23163,3.97482)
 (15.11901,14.96691)
 (32.67648,32.36862)
 (2.08238,1.92409)
 (33.46921,33.21224)
 (22.59151,22.52659)
 (42.37024,42.60059)
 (1.24149,1.47601)
 (3.44607,3.23903)
 (21.45549,21.63039)
 (27.41250,27.48804)
 (41.97952,41.72896)
 (47.63400,47.54626)
 (43.99194,43.87614)
 (22.89227,22.92286)
 (8.00572,7.71471)
 (35.51702,35.70482)
 (8.93071,8.65630)
 (5.84366,5.54502)
 (23.83463,23.83293)
 (32.16468,31.89043)
 (13.64213,13.61864)
 (17.41878,17.56812)
 (3.84305,3.63551)
 (38.90229,38.86789)
 (9.57916,9.75988)
 (32.46486,29.39314)
 (26.63094,26.42055)
 (28.15415,27.99069)
 (7.40618,7.61851)
 (12.95791,12.93052)
 (19.71853,20.00916)
 (30.27415,30.73965)
 (35.23383,34.89648)
 (1.37787,1.48297)
 (0.70486,0.36799)
 (13.63286,13.58245)
 (23.24215,23.21247)
 (14.01665,14.28405)
 (7.21252,7.23087)
 (6.05413,6.19483)
 (33.73248,33.67901)
 (17.24092,17.08056)
 (19.25250,19.17520)
 (6.42456,6.54049)
 (13.57734,13.53623)
 (1.97325,2.30341)
 (8.86337,8.93802)
 (0.42637,0.76926)
 (25.35804,25.54887)
 (18.15328,18.35765)
 (11.99103,11.71035)
 (21.85163,21.82553)
 (27.48223,28.10318)
 (19.58582,23.41802)
 (6.00516,6.03609)
 (1.95943,2.83892)
 (36.04578,37.09743)
 (2.29960,2.17742)
 (0.44870,0.26606)
 (13.71719,14.41042)
 (16.39900,16.13952)
 (37.28839,37.33289)
 (29.01250,28.85576)
 (17.28925,17.10999)
 (50.60226,51.31755)
 (7.67469,9.10515)
 (21.36586,21.04290)
 (30.27644,30.21826)
 (20.42066,20.48957)
 (12.39896,12.49165)
 (18.51739,18.07049)
 (45.27872,45.13180)
 (22.06419,22.07316)
 (13.52259,13.44292)
 (36.46746,36.45229)
 (0.93641,0.82996)
 (15.58042,15.96238)
 (55.48829,55.17756)
 (2.62355,2.73688)
 (4.34315,4.32537)
 (29.89379,30.14048)
 (52.25879,52.20555)
 (73.75631,73.43571)
 (2.80566,3.51974)
 (2.15936,2.11545)
 (41.40630,41.17599)
 (17.93680,18.02291)
 (16.26113,16.02338)
 (30.71017,30.50440)
 (5.36756,5.26745)
 (17.36097,17.20879)
 (60.66192,60.44800)
 (5.45761,5.75006)
 (14.23767,14.34864)
 (0.31058,0.11577)
 (6.80796,6.40941)
 (39.94650,40.01334)
 (7.76530,7.79241)
 (12.46604,12.22187)
 (12.15174,12.42121)
 (2.69611,2.38233)
 (30.80567,31.37002)
 (13.29931,13.02366)
 (11.95262,11.96036)
 (29.58662,29.69527)
 (53.39896,60.86144)
 (8.23849,8.84850)
 (19.45836,19.30372)
 (14.83174,14.57983)
 (8.99570,8.91048)
 (18.82729,18.86063)
 (2.09608,1.83715)
 (3.05333,3.15268)
 (32.91180,33.08755)
 (26.89172,27.12613)
 (28.85690,28.97783)
 (16.42227,16.42931)
 (7.46810,7.71213)
 (4.86590,4.72596)
 (39.30622,38.89814)
 (27.22628,27.07014)
 (9.48475,9.65367)
 (70.64721,70.58369)
 (28.76156,28.75909)
 (17.82897,18.08998)
 (10.76146,11.14124)
 (1.63912,1.40985)
 (33.08278,32.90777)
 (48.24931,48.02389)
 (21.33587,21.41932)
 (19.71621,19.48349)
 (14.16415,14.17362)
 (12.76157,12.94833)
 (6.73440,7.13141)
 (36.46469,36.49031)
 (7.98785,8.37075)
 (11.57024,12.09908)
 (17.64439,17.67734)
 (4.93983,5.00753)
 (35.34523,35.05235)
 (6.27007,6.18307)
 (0.13716,0.15876)
 (30.40403,30.19268)
 (3.62400,3.64984)
 (22.29452,22.44259)
 (5.51869,5.26420)
 (8.85155,8.69040)
 (40.29949,31.58496)
 (9.43377,9.08395)
 (6.08418,5.98760)
 (21.77283,21.66958)
 (4.99368,4.41165)
 (45.92191,43.19888)
 (5.86736,6.10968)
 (11.74432,11.54419)
 (24.61121,25.16978)
 (4.22601,2.64792)
 (54.03383,54.75772)
 (42.22046,42.15826)
 (6.86110,6.85923)
 (5.64475,6.06886)
 (0.53505,0.22329)
 (15.19018,15.33696)
 (35.86574,35.91336)
 (39.39872,39.57553)
 (4.81997,4.97211)
 (63.47731,63.43019)
 (18.28147,18.53105)
 (25.13994,25.07914)
 (14.11913,14.26076)
 (16.36815,16.29848)
 (9.25507,9.28934)
 (9.57397,9.54324)
 (31.51800,31.72531)
 (14.57864,14.52365)
 (25.62035,25.92087)
 (23.89938,23.93499)
 (46.57344,46.41150)
 (9.10698,8.68338)
 (8.95972,9.06050)
 (12.93438,13.02755)
 (2.39177,0.40142)
 (18.03175,18.03572)
 (48.70109,48.67097)
 (29.72432,29.67232)
 (42.61505,42.46288)
 (30.17209,29.99855)
 (12.61181,12.68481)
 (42.74797,42.78752)
 (26.12837,26.07961)
 (14.53531,13.10790)
 (25.26814,25.43792)
 (30.39089,30.19739)
 (68.83671,68.74103)
 (29.40361,29.15954)
 (22.34565,22.21115)
 (19.46059,19.20103)
 (12.27341,12.03334)
 (8.91477,9.20663)
 (5.34212,5.50534)
 (11.88211,11.99822)
 (20.45294,19.42938)
 (51.10611,51.12622)
 (9.62230,9.75518)
 (43.66342,44.11776)
 (23.21917,23.24984)
 (6.97602,6.80639)
 (25.40567,25.46304)
 (9.68869,9.73896)
 (17.50700,18.36770)
 (15.65081,15.67757)
 (2.99018,3.40905)
 (6.18481,5.87180)
 (27.47901,27.22911)
 (12.71724,12.76846)
 (15.23491,15.24257)
 (33.26760,33.41782)
 (60.24513,59.46121)
 (1.04712,1.49652)
 (45.29743,45.14708)
 (8.45361,9.29505)
 (1.49130,1.17064)
 (15.66234,15.55917)
 (0.97849,0.63132)
 (8.12562,7.97003)
 (20.96144,21.16694)
 (30.22108,30.56734)
 (16.20333,16.51413)
 (17.68583,17.68380)
 (23.51990,23.37347)
 (5.38001,6.18876)
 (8.07890,8.03447)
 (7.04148,7.00208)
 (18.14440,18.23305)
 (7.07202,7.37840)
 (18.23396,17.34587)
 (14.34168,14.65835)
 (49.14082,49.23575)
 (3.80515,3.96529)
 (30.49823,30.28429)
 (12.44581,12.24690)
 (0.24812,0.08390)
 (10.29466,11.02775)
 (14.14124,50.22253)
 (77.91252,78.13385)
 (16.55677,16.68906)
 (3.19197,3.00266)
 (43.27869,43.47152)
 (19.51360,19.71600)
 (3.74660,2.44101)
 (6.76933,7.15717)
 (9.04512,9.05988)
 (7.17424,7.25838)
 (0.30130,0.16858)
 (30.41188,30.62988)
 (29.08321,27.95158)
 (10.06960,10.70895)
 (3.98290,3.83067)
 (11.48079,11.85982)
 (7.40508,7.11145)
 (37.17503,37.36768)
 (30.76106,31.36138)
 (34.37393,34.30141)
 (5.92971,5.59562)
 (24.38105,24.57478)
 (37.96497,38.31995)
 (39.72350,40.03300)
 (0.63547,1.01539)
 (13.03165,12.39932)
 (0.58820,0.20464)
 (24.58636,24.48998)
 (5.82731,5.52068)
 (17.83802,17.85336)
 (19.11043,19.17642)
 (7.76573,6.64445)
 (29.95956,30.33553)
 (10.98643,11.10366)
 (40.06726,39.60816)
 (20.06410,20.00603)
 (8.29953,8.00100)
 (9.29498,9.35989)
 (16.43861,16.37875)
 (6.53278,6.03667)
 (48.42221,38.12609)
 (26.11697,26.06664)
 (6.71313,7.42263)
 (10.72401,10.94659)
 (2.17398,2.05675)
 (16.41509,16.40760)
 (34.68328,34.83371)
 (9.95599,10.11507)
 (27.61903,27.87208)
 (9.78354,9.80527)
 (32.81698,32.59596)
 (9.39014,9.51302)
 (43.06317,43.14677)
 (33.74021,33.68194)
 (27.36468,26.23183)
 (3.01721,2.76496)
 (21.14727,21.11561)
 (16.46399,16.38174)
 (11.09906,11.19590)
 (32.04593,32.11921)
 (6.32512,6.08566)
 (27.82743,27.72763)
 (27.39174,27.38809)
 (21.11720,21.28738)
 (0.18600,0.32801)
 (8.77356,9.15666)
 (52.09479,52.74701)
 (18.14753,19.16236)
 (38.60941,38.49591)
 (39.61283,39.60359)
 (6.97997,7.02731)
 (16.94176,17.33024)
 (9.31978,9.57818)
 (8.23086,8.12456)
 (14.09681,13.98087)
 (31.70058,31.72666)
 (36.68709,36.70029)
 (10.59631,10.70259)
 (6.87902,6.93629)
 (30.94071,33.56278)
 (30.81458,30.83134)
 (6.95656,7.10342)
 (20.98013,20.77931)
 (4.12783,4.31258)
 (1.78030,1.72600)
 (19.28335,19.13045)
 (12.54616,12.56499)
 (12.99980,12.25419)
 (8.89677,8.75159)
 (5.44749,5.49836)
 (16.13881,15.81736)
 (26.68036,26.77975)
 (17.96453,18.43456)
 (45.67908,45.86139)
 (5.25743,5.10918)
 (0.28226,1.07002)
 (16.75654,16.08683)
 (30.79646,30.91967)
 (39.52825,39.66514)
 (41.97163,42.00570)
 (1.71416,1.48588)
 (19.66546,19.61503)
 (30.80339,25.82306)
 (56.96837,56.48803)
 (4.37650,4.89401)
 (2.90664,2.99805)
 (37.07541,36.99751)
 (8.19227,2.28542)
 (0.05039,0.57144)
 (14.92996,14.82453)
 (11.28508,11.29419)
 (27.17436,27.23470)
 (9.29533,9.14955)
 (37.18596,37.37041)
 (0.13251,0.75628)
 (6.06467,6.10605)
 (4.20051,4.08282)
 (4.12026,3.34536)
 (10.00785,10.13683)
 (0.35050,0.54903)
 (13.13311,12.69066)
 (8.09005,7.65495)
 (44.52937,48.53933)
 (11.08414,11.15405)
 (11.28764,11.26796)
 (11.12887,11.29805)
 (9.36864,9.69594)
 (20.17111,20.09176)
 (6.10509,2.16315)
 (22.01894,21.91546)
 (52.52925,52.45841)
 (31.10831,31.08781)
 (22.04612,22.09758)
 (5.21960,6.02387)
 (30.17123,30.14518)
 (18.05948,18.35159)
 (21.29212,20.76811)
 (2.88617,2.76645)
 (11.93914,12.14024)
 (10.09223,9.80496)
 (1.90909,2.13209)
 (12.19566,12.17730)
 (22.93371,22.82308)
 (6.59422,6.53114)
 (13.05042,12.33907)
 (37.54846,37.24689)
 (23.36724,23.02279)
 (19.15781,22.38008)
 (21.35297,21.24142)
 (23.11559,23.33798)
 (37.55554,38.29748)
 (25.71619,26.58894)
 (7.23373,7.26869)
 (17.32880,17.70284)
 (5.93745,5.70995)
 (3.60191,3.44015)
 (36.56037,36.55112)
 (0.29354,0.15113)
 (12.92172,12.73160)
 (20.45058,20.75217)
 (14.30830,14.44364)
 (1.90671,1.96194)
 (13.42705,13.70688)
 (8.39351,7.90823)
 (57.98667,58.11822)
 (15.57525,15.36426)
 (18.54509,18.84723)
 (30.02590,30.15880)
 (21.82864,21.05753)
 (5.56086,5.68143)
 (27.89144,28.08202)
 (7.55275,7.71011)
 (0.10152,9.90950)
 (48.74969,48.76413)
 (6.98097,6.93993)
 (4.08332,3.64335)
 (7.24100,6.97250)
 (10.67488,10.55994)
 (29.34933,29.78435)
 (16.13086,10.78902)
 (25.90654,26.16702)
 (49.48841,49.94137)
 (2.06437,1.96538)
 (22.01351,22.16836)
 (14.75941,14.61471)
 (4.08046,4.21960)
 (32.93571,32.81908)
 (28.01130,27.55270)
 (51.52665,50.51430)
 (6.27790,6.23676)
 (20.44790,20.18067)
 (9.51638,9.60025)
 (24.05318,24.05930)
 (50.55794,51.10032)
 (0.37270,0.10707)
 (24.74403,24.84430)
 (26.71607,26.47521)
 (26.26484,25.95790)
 (55.20867,55.23330)
 (21.70222,21.61324)
 (4.10162,4.06138)
 (1.17717,1.59951)
 (0.02444,0.24657)
 (11.78000,11.39955)
 (11.47501,11.23057)
 (3.24042,3.40631)
 (8.94533,9.58899)
 (19.61420,19.77427)
 (44.48233,43.67596)
 (33.81500,33.72559)
 (32.54923,32.39659)
 (5.66925,2.62627)
 (32.62699,32.46560)
 (0.00514,0.43203)
 (28.26364,27.19940)
 (1.49265,1.17016)
 (11.40950,11.66694)
 (37.23196,37.27871)
 (19.12434,18.95527)
 (12.59042,13.08099)
 (1.45201,1.59363)
 (30.12911,30.22954)
 (8.87826,9.27940)
 (2.21689,15.07911)
 (43.32259,42.56894)
 (29.03547,28.97956)
 (5.37914,5.40454)
 (3.75496,3.76971)
 (5.53848,5.22281)
 (9.58762,9.64135)
 (15.35217,15.08635)
 (5.51131,5.61933)
 (14.07462,14.14004)
 (9.41666,9.48910)
 (32.40862,31.92120)
 (26.97121,26.90749)
 (51.54241,51.60052)
 (28.95683,29.13493)
 (5.50177,5.92204)
 (4.89661,4.70015)
 (6.51650,6.47216)
 (20.45779,19.94283)
 (5.84094,6.05799)
 (7.88530,8.07219)
 (22.88631,22.79892)
 (16.81904,16.80523)
 (11.30291,11.55137)
 (0.50091,0.05825)
 (9.74036,9.88628)
 (32.19925,31.86895)
 (1.06874,1.05750)
 (32.83505,37.10180)
 (8.80502,9.12526)
 (16.66150,16.71159)
 (29.47883,29.35675)
 (21.69778,21.47792)
 (39.05814,39.14019)
 (43.83251,43.39795)
 (0.48590,0.08201)
 (26.72311,26.22306)
 (18.51280,18.28831)
 (0.08264,10.79333)
 (21.62663,21.82650)
 (16.73226,17.11502)
 (30.28173,30.34439)
 (35.67462,35.75527)
 (10.78271,10.38744)
 (14.38860,13.86888)
 (12.93959,12.69524)
 (26.77816,26.60935)
 (9.63068,9.74351)
 (40.67665,40.66406)
 (8.08074,8.18925)
 (0.61515,0.15934)
 (14.88558,15.11728)
 (1.80502,1.67216)
 (21.42620,21.39931)
 (10.10177,9.99502)
 (27.40755,27.33864)
 (3.47876,3.55275)
 (15.20642,15.13627)
 (12.00622,11.91332)
 (9.90234,9.10271)
 (1.53284,1.07988)
 (26.32910,26.84528)
 (6.15405,5.63703)
 (0.38161,0.45949)
 (16.08279,16.58078)
 (1.82156,1.78973)
 (54.42328,43.75268)
 (11.09597,7.27757)
 (46.81949,46.79461)
 (7.71594,7.81324)
 (6.80964,6.49477)
 (15.64863,15.41487)
 (15.66215,15.58450)
 (24.48411,24.17560)
 (32.94357,32.75832)
 (1.51660,1.72900)
 (32.63827,32.86781)
 (9.87293,9.90987)
 (20.27190,19.72559)
 (6.54554,7.03678)
 (8.47445,8.73713)
 (55.18473,55.28815)
 (3.07157,2.87029)
 (17.50315,17.33964)
 (41.01025,40.87734)
 (0.57436,0.21751)
 (24.96926,25.19489)
 (7.40460,7.09436)
 (9.89315,9.78152)
 (18.79450,18.52195)
 (11.46299,11.13494)
 (9.22335,9.05312)
 (24.59321,24.72448)
 (27.42302,27.34225)
 (10.80885,10.43581)
 (2.52426,2.40850)
 (9.61089,9.97508)
 (25.68498,25.65966)
 (6.95793,12.28150)
 (20.12952,20.35972)
 (16.08305,16.02403)
 (40.77334,40.85337)
 (14.29211,13.71838)
 (6.26603,5.95554)
 (17.42147,17.44867)
 (0.34684,0.25283)
 (6.31160,6.24831)
 (10.96420,10.33817)
 (14.01937,14.25711)
 (2.49556,2.63308)
 (10.47213,10.57853)
 (20.99100,20.80641)
 (20.40666,20.56483)
 (3.25211,3.30035)
 (2.57778,2.70960)
 (0.57182,0.58565)
 (1.07313,0.56711)
 (7.91016,7.69572)
 (36.40409,36.35638)
 (45.28668,45.31360)
 (2.43875,2.11420)
 (61.28864,61.10935)
 (13.58858,14.03053)
 (2.04954,1.49380)
 (43.67586,43.75479)
 (61.65672,61.21988)
 (4.75015,4.93696)
 (12.32608,13.05713)
 (63.73185,63.96756)
 (21.32162,21.44384)
 (29.72688,29.55413)
 (9.92721,9.89321)
 (15.86962,15.67695)
 (21.63958,21.71775)
 (1.66160,1.11515)
 (13.16235,13.04699)
 (9.52714,9.41232)
 (1.07021,0.66027)
 (1.38919,1.60480)
 (30.07800,30.07981)
 (40.15872,39.61739)
 (37.66718,37.71191)
 (20.47852,23.67868)
 (20.34245,20.45520)
 (29.73367,29.61575)
 (77.51278,77.11549)
 (20.70214,20.88386)
 (15.39572,15.33096)
 (30.09573,29.56270)
 (40.84041,41.62930)
 (33.15793,33.22802)
 (59.37973,59.93118)
 (19.63800,13.61090)
 (7.09690,7.06009)
 (0.77832,0.98864)
 (1.47808,1.69531)
 (40.47159,40.48415)
 (9.52093,9.69420)
 (31.69198,31.68607)
 (10.58540,10.96186)
 (9.43754,9.17762)
 (22.75547,22.64210)
 (10.76436,10.71048)
 (58.15322,57.49509)
 (57.99638,59.06971)
 (37.17212,36.73626)
 (57.69402,57.70736)
 (11.67993,12.09665)
 (29.93900,29.84614)
 (0.96238,0.42021)
 (35.55652,35.15942)
 (36.92006,36.58149)
 (1.83779,1.03794)
 (2.31597,2.30141)
 (3.49979,3.59977)
 (22.79914,22.95761)
 (35.52260,35.31710)
 (14.19352,14.71609)
 (40.51500,39.87365)
 (25.49192,25.34408)
 (1.62519,1.61649)
 (7.48967,7.92230)
 (8.44714,8.36815)
 (3.02690,2.87926)
 (6.55870,6.53803)
 (16.45113,16.31149)
 (2.87141,2.89210)
 (0.91106,1.60238)
 (5.22645,5.17717)
 (15.20872,15.16278)
 (10.13175,10.36828)
 (6.46764,6.44886)
 (3.09686,3.31833)
 (14.58220,14.30525)
 (1.29034,0.40084)
 (9.09389,8.95788)
 (15.52869,15.31522)
 (0.47289,0.42877)
 (9.99049,9.70410)
 (57.21026,56.76212)
 (11.31104,11.58928)
 (16.61131,16.84456)
 (31.70456,31.92068)
 (38.32825,38.36430)
 (25.68539,25.68978)
 (0.77451,0.98334)
 (5.04185,5.11646)
 (28.45664,24.82788)
 (18.66771,18.62673)
 (2.11217,4.26776)
 (10.90997,11.03856)
 (14.47056,14.34515)
 (1.30708,1.55804)
 (6.44812,6.16040)
 (2.54494,1.99440)
 (32.64318,32.52207)
 (43.33369,43.17508)
 (44.62144,44.63801)
 (15.55819,15.77815)
 (2.94139,2.77787)
 (20.84090,20.69613)
 (47.15947,47.26551)
 (37.87634,37.80602)
 (9.91381,10.00017)
 (40.14594,40.36466)
 (12.87667,12.82082)
 (9.16450,9.66968)
 (41.27062,41.43820)
 (57.75519,58.16102)
 (16.57024,16.42924)
 (8.94883,9.12671)
 (1.11975,1.76969)
 (20.08241,20.40209)
 (17.40029,17.50170)
 (28.61876,39.18980)
 (21.35940,21.90020)
 (2.25573,2.23143)
 (5.09838,4.41287)
 (2.29951,2.15616)
 (3.22211,2.98462)
 (23.90463,24.28719)
 (39.58290,39.34545)
 (29.15963,29.45623)
 (6.97160,6.79747)
 (32.82578,29.39500)
 (7.31053,7.48725)
 (10.96251,11.11516)
 (13.21220,12.90300)
 (11.70780,11.73790)
 (27.39659,27.82554)
 (11.55589,16.37653)
 (17.04486,16.67519)
 (30.23022,30.76569)
 (5.23766,5.20757)
 (1.54494,1.89151)
 (41.73993,41.68140)
 (46.71687,46.58797)
 (6.64166,6.13176)
 (1.24673,1.28082)
 (15.61908,15.61289)
 (6.77712,6.50480)
 (16.07282,16.03130)
 (2.85665,6.46898)
 (6.29149,6.36634)
 (60.05012,59.84285)
 (16.94984,17.10610)
 (21.76881,21.79241)
 (17.91741,17.61159)
 (11.88549,10.76068)
 (23.83287,24.20068)
 (31.30415,31.37728)
 (11.73878,11.95891)
 (3.24024,3.15642)
 (37.19022,37.33582)
 (24.66198,24.73751)
 (11.14062,10.39550)
 (22.22708,22.10312)
 (10.40839,14.37628)
 (24.10890,24.04565)
 (25.59417,25.67568)
 (30.69457,30.30243)
 (34.27620,30.40364)
 (38.07264,37.80912)
 (40.10949,40.60242)
 (3.98806,4.23853)
 (13.66647,14.06116)
 (3.61663,2.74767)
 (0.40728,0.28466)
 (1.72623,1.59961)
 (28.73109,29.00454)
 (43.88852,43.63334)
 (8.85504,8.87880)
 (16.68227,16.95148)
 (15.52632,14.89834)
 (22.09101,22.20632)
 (8.89377,9.19009)
 (14.28632,14.61955)
 (1.26837,1.48080)
 (32.83436,27.25325)
 (3.96215,4.03967)
 (7.02049,7.60004)
 (3.71039,4.05686)
 (3.79781,3.87973)
 (53.36445,53.13191)
 (30.65468,30.42887)
 (22.43960,22.47039)
 (2.34100,2.12222)
 (27.15387,27.38066)
 (0.99169,1.13023)
 (32.27062,32.13729)
 (42.66682,42.75914)
 (0.87809,1.01438)
 (33.48170,33.19607)
 (34.92116,34.84253)
 (5.79456,5.76641)
 (20.41048,20.23630)
 (55.37479,55.14526)
 (1.46024,2.29889)
 (17.69937,17.88309)
 (10.65459,10.52509)
 (39.76497,40.22058)
 (15.49955,15.39347)
 (16.62291,16.63889)
 (28.23589,28.11592)
 (20.56207,20.18270)
 (23.26494,23.28060)
 (2.64460,2.23404)
 (19.38213,18.99831)
 (5.11199,4.80757)
 (8.28025,8.47742)
 (9.18655,9.35720)
 (24.20060,24.66465)
 (7.98530,7.98303)
 (6.03068,6.03072)
 (13.53747,14.00073)
 (22.60441,22.60600)
 (6.53117,6.96855)
 (10.08062,10.15171)
 (9.82259,9.78847)
 (19.58927,19.52719)
 (27.49860,27.38698)
 (40.31299,40.23095)
 (24.71198,24.70937)
 (20.64142,20.70070)
 (10.61688,10.82055)
 (16.63054,16.69941)
 (12.90651,12.58987)
 (23.41835,23.67533)
 (23.99968,23.86072)
 (37.44056,37.53591)
 (14.77406,14.64027)
 (0.18260,0.17370)
 (4.83448,5.73684)
 (12.04014,11.94432)
 (7.13935,6.79519)
 (28.68414,28.35486)
 (6.94959,6.90583)
 (23.42585,23.24187)
 (32.51661,32.83641)
 (31.09374,31.60338)
 (27.30634,27.31099)
 (0.15048,0.08978)
 (15.66299,15.15639)
 (14.77837,14.71197)
 (15.10771,15.08207)
 (41.07077,41.19533)
 (10.48396,10.49118)
 (8.01064,7.84774)
 (19.35165,19.34012)
 (7.07096,6.95538)
 (44.27699,44.27856)
 (8.17546,8.01805)
 (37.61107,37.59478)
 (24.92902,24.93990)
 (59.12661,58.68739)
 (24.47111,24.39131)
 (11.09844,11.04820)
 (2.07104,1.66829)
 (17.25276,17.61429)
 (48.44741,48.73788)
 (47.98832,47.84283)
 (28.21841,28.73775)
 (63.10292,62.98651)
 (39.60289,39.62947)
 (17.83963,17.59870)
 (8.35151,8.42797)
 (6.14950,6.43576)
 (7.92161,7.71088)
 (4.83891,5.04186)
 (45.25226,45.20121)
 (17.02178,17.18415)
 (4.14078,4.01050)
 (35.25610,35.94842)
 (25.36568,25.27233)
 (21.89911,21.72346)
 (41.98707,41.84887)
 (28.62354,28.22124)
 (10.39246,10.12085)
 (10.66838,10.28649)
 (39.06018,38.99331)
 (4.86878,4.84201)
 (30.64805,30.59230)
 (14.08702,14.07407)
 (6.95346,6.87819)
 (1.83309,2.10422)
 (6.82963,1.41952)
 (7.63130,7.02694)
 (6.49116,6.75702)
 (8.02607,7.98051)
 (4.15852,4.03293)
 (32.91368,32.64887)
 (62.95049,62.83304)
 (1.26506,1.01609)
 (4.18168,3.76104)
 (16.40499,16.56303)
 (14.98109,14.37526)
 (28.24844,23.06261)
 (14.91915,25.17833)
 (18.66053,19.01618)
 (8.18261,8.20985)
 (25.53496,25.47470)
 (30.21067,29.97033)
 (44.57343,43.90854)
 (3.31530,3.22822)
 (38.29590,38.06536)
 (41.15329,40.94818)
 (43.68782,43.75535)
 (54.37570,54.38895)
 (7.86753,7.89838)
 (10.60678,10.58833)
 (21.31318,21.46617)
 (59.67293,44.96489)
 (10.32635,9.86444)
 (48.45409,48.63779)
 (40.64023,40.98146)
 (3.17535,3.43483)
 (24.85903,25.12721)
 (18.48531,18.62590)
 (35.61013,34.99711)
 (4.89297,4.92863)
 (10.81434,11.17407)
 (2.74131,2.71433)
 (34.06091,34.54747)
 (10.55641,10.10073)
 (16.34785,16.51731)
 (24.62144,24.55079)
 (54.63488,54.33850)
 (17.52609,17.25972)
 (31.96319,32.07711)
 (19.76617,19.98908)
 (43.47004,43.03922)
 (37.24747,37.08581)
 (12.39756,12.88904)
 (0.64930,0.50605)
 (7.42710,6.94371)
 (51.71164,51.51473)
 (58.97688,58.85080)
 (56.58222,56.23645)
 (13.40626,13.12735)
 (11.59365,11.44616)
 (0.01256,0.74290)
 (50.82810,51.33463)
 (15.93209,15.59426)
 (10.89239,11.03915)
 (34.70658,35.83437)
 (44.64104,44.33900)
 (23.73506,23.88454)
 (10.15955,10.03025)
 (16.22088,16.43242)
 (3.67064,3.90487)
 (37.42987,37.01266)
 (41.84373,41.87338)
 (70.20627,70.45465)
 (56.08700,56.02362)
 (13.79070,13.90090)
 (11.19371,10.94044)
 (2.65647,2.58817)
 (7.13461,6.51885)
 (7.10919,8.09085)
 (24.65669,24.76146)
 (28.17772,28.00143)
 (7.65053,7.46738)
 (63.65612,63.81610)
 (13.44529,13.17673)
 (0.08138,0.70511)
 (8.16940,8.28410)
 (14.58974,14.14074)
 (12.99136,12.93897)
 (22.71775,22.75487)
 (11.00607,10.49524)
 (34.32011,34.28593)
 (13.47339,13.76027)
};
\addlegendentry{\textsf{Local} ($r=0.995$, $\rho=0.995$)}
\end{axis}
\end{tikzpicture}
\caption{Actual \textbf{encoder velocity estimation error} (\ref{eq:vel_loss}) versus \textbf{velocity disagreement} (\ref{eq:velocity_disagreement}) under two different
anomalies with strength 0.6 (randomly chosen $n=1000$ per anomaly type, Pearson $r$, Spearman $\rho$) on $\mathcal{D}^{\mathrm{test}}$.}
\label{fig:velocity_scatter}
\end{figure}

%% file: sections/6_Conclucions.tex
\section{Conclusions}
\label{sec:conclusions}
This paper investigated conditional generative modeling for assessing Doppler RaDAR reliability in railway speed-over-ground estimation. Three complementary indicators were evaluated: the spectral anomaly score (SAS), conditional anomaly score (CAS), and absolute velocity disagreement (AVD). The learned spectral model showed clear sensitivity to velocity deviations. Under controlled anomaly conditions, SAS and CAS outperformed conventional spectral indicators in most conditions, particularly for spectral shifts, additional peaks, and local distortions. AVD showed a strong correlation with actual estimation errors for the investigated extra-peak and local anomalies. These findings highlight the potential of conditional generative spectral modeling for assessing spectral plausibility and velocity estimation reliability.
\section{Acknowledgment}
The research leading to these results is funded by the German Federal Ministry for Economic Affairs and Energy within the project “CONTROL – Controlling Risk for Highly Automated Transportation Systems Operating in Complex Open Environments". The authors would like to thank the consortium for the successful cooperation.